\documentclass[useAMS, usenatbib, usegraphicx]{mn2e} 
\usepackage{deluxetable} 

\title[SPH Simulations of BH Accretion] 
{SPH Simulations of Black Hole Accretion: A Step to Model Black Hole Feedback in Galaxies}

\author[P. Barai et al.] 
{Paramita Barai\thanks{E-mail: barai@physics.unlv.edu}, 
Daniel Proga, and Kentaro Nagamine\\ 
Department of Physics \& Astronomy, 
University of Nevada, Las Vegas, 4505 S. Maryland Parkway, Box 454002, \\ 
Las Vegas, NV, 89154-4002, USA.}

\begin{document} 



\maketitle

\label{firstpage}

\begin{abstract}

We test how accurately the smoothed particle hydrodynamics (SPH) numerical technique 
can follow spherically-symmetric Bondi accretion. 
Using the 3D SPH code GADGET-3, we perform simulations of gas accretion
onto a central supermassive black hole (SMBH) of mass $10^8 M_{\odot}$ within the radial range of $0.1 - 200$ pc. 
We carry out simulations without and with radiative heating by a central X-ray corona and radiative cooling. 
For an adiabatic case, the radial profiles of hydrodynamical properties match the Bondi solution, 
except near the inner and outer radius of the computational domain. 
The deviation from the Bondi solution close to the inner radius is caused by 
the combination of numerical resolution, artificial viscosity, and our inner boundary condition. 
Near the outer radius ($\leq 200$ pc), 
we observe either an outflow or development of a non-spherical inflow 
unless the outer boundary conditions are very stringently implemented. 
Despite these issues related to the boundary conditions, 
we find that adiabatic Bondi accretion can be reproduced for durations of a few dynamical times at the Bondi radius, 
and for longer times if the outer radius is increased.   
In particular, the mass inflow rate at the inner boundary, which we measure, is within $3 - 4 \%$ of the Bondi accretion rate. 
With radiative heating and cooling included, the spherically accreting gas 
takes a longer time to reach a steady-state than the adiabatic Bondi accretion runs, 
and in some cases does not reach a steady-state even within several hundred dynamical times. 
We find that artificial viscosity causes excessive heating near the inner radius, 
making the thermal properties of the gas inconsistent with a physical solution. 
This overheating occurs typically only in the supersonic part of the flow, 
so that it does not affect the mass accretion rate. 
We see that increasing the X-ray luminosity produces a lower central mass inflow rate, 
implying that feedback due to radiative heating is operational in our simulations. 
With a sufficiently high X-ray luminosity, the inflowing gas is radiatively heated up, and an outflow develops. 
We conclude that the SPH simulations can capture the gas dynamics needed to study radiative feedback 
provided artificial viscosity alters only highly supersonic part of the inflow. 

\end{abstract}



\section{Introduction} 
\label{sec-intro} 

Accretion of gas onto supermassive black holes (SMBHs) at the centers of active galaxies 
and the resulting feedback from them have strongly influenced the formation and evolution of galaxies 
\citep[e.g.,][]{salpeter64, lynden-Bell69, rees84, kormendy95, Ciotti97, richstone98, 
silk98, Blandford99, Fabian99, ferrarese00, King03, granato04, best07, Barai08, Germain09}. 
It likely causes observational trends such as the central SMBH--host galaxy correlations 
\citep[e.g.,][]{magorrian98, gebhardt00}. 
The concordance model of structure formation based on $\Lambda$ cold dark matter cosmology 
invokes feedback from SMBHs as a crucial ingredient to self-regulate galaxy and SMBH growth, 
as have been studied in numerical hydrodynamic simulations 
\citep[e.g.,][]{Ciotti01, DiMatteo05, Robertson06, Thacker06, Li07, AntonuccioDelogu10, Ostriker10, Novak10},   
as well as in semi-analytic models of galaxy formation 
\citep[e.g.,][]{kauffmann00, Bower06, Croton06, malbon07, Somerville08}. 

In galaxy formation simulations that can resolve kpc to 100's of pc scales, 
the accretion flow onto a SMBH occurs in unresolved scales, 
and a subgrid prescription is needed to model SMBH accretion. 
The Bondi-Hoyle-Lyttleton parameterization \citep{Hoyle39, Bondi44, Bondi52}, 
relating the SMBH mass accretion rate to the resolved larger scale properties of the ambient gas, 
is often assumed in an ad hoc manner. 
Such an assumption has been made in studies of 
isolated systems and mergers \citep[e.g.,][]{SDH05, Johansson09b, Lusso10} 
and cosmological simulations \citep[e.g.,][]{Sijacki07, DiMatteo08, Booth09, Dubois10}. 
The lack of resolution introduces uncertainties 
in estimating the gas density and temperature near the BH and hence the BH accretion rate.   
As a numerical correction,  
the accretion rate inferred from simulations is multiplied by a factor of the order of $100$   
\citep[e.g.,][]{SDH05, Sijacki07, DiMatteo08, Booth09, Sijacki09}.       
The ISM gas density and temperature are estimated 
by smoothing on the scale of the computational resolution at the location of the BH. 
This results in artificially low densities and high temperatures compared to the case where the 
multiphase structure of the ISM could be spatially resolved on the scale of the Bondi radius. 
As a compensation the Bondi accretion rate is multiplied by $100$ \citep{Johansson09a} or $300$ \citep{Khalatyan08}. 
\citet{Booth09} used a density dependent multiplication factor. 
Most of these simulations were performed using the GADGET code \citep{Springel05}. 


GADGET is a smoothed particle hydrodynamics \citep[SPH,][]{Gingold77, Lucy77, Monaghan92} code. 
SPH uses a Lagrangian method for modeling fluid dynamics, 
which can handle regions of higher density with higher resolution, and can simulate large dynamic ranges. 
SPH is widely used in structure formation studies on cosmological scales. 
The GADGET code has been demonstrated to pass several tests in the original publication \citep{Springel05}: 
standard Sod shock-tube, self-gravitational adiabatic collapse, isothermal collapse, 
interaction of a strong shock with a dense gas cloud, 
dark matter halo mass function and clustering, and formation of a rich galaxy cluster. 
Other studies have also used GADGET for test problems in various context: 
gravitational collapse and fragmentation of molecular cloud cores \citep{Arreaga-Garcia07}, 
bubble growth during hydrogen reionization \citep{Zahn07}, 
detection of shock waves in cosmological simulations \citep{Pfrommer06}, 
rotation curves of disk galaxies \citep{Kapferer06}, 
turbulent gas motions in the intracluster medium \citep{Dolag05, Vazza06}, 
and baryon physics in galaxy clusters \citep{Borgani06}. 
However, in spite of several studies of SMBH feedback assuming Bondi accretion using GADGET, 
a rigorous test of the simple Bondi accretion, or the presentation of results from such a test, 
has been absent from the literature. 

Here, we present the results of our test simulations of SMBH accretion using the GADGET code. 
We focus on purely spherically symmetric accretion neglecting any motion of the BH with respect to the ambient gas. 
The question that we want to answer is, ``What is required of an SPH code to reproduce the Bondi problem?''.  
We check how well the code can reproduce the Bondi mass accretion solution, and 
investigate the dependence of the central mass accretion rate on parameter values, length and time scales. 
We also attempt to study one of the modes of AGN feedback: radiation heating. 

A rigorous estimate of BH accretion rate requires a spatial resolution of the order of a sonic radius, 
or at least of the order of the Bondi radius, 
which is not possible in current cosmological simulations. 
An alternative approach would be to perform simulations by reducing the size of the computational volume. 
We follow this alternative approach, and in our current study we resolve the Bondi and sonic radii. 
Effectively we test if it is possible to refine the subgrid model of SMBH accretion 
by measuring the BH accretion rate in higher-resolution simulations using the same code that is used for cosmological simulations. 

Similar work has been done using other numerical techniques, to which the results of our SPH-based study can be compared. 
A contemporary work by \citet{KPN09} performed two and three dimensional radiation hydrodynamic simulations 
of relatively large-scale outflows ($\sim 0.01 - 10$ pc) from SMBHs using the 
Eulerian code ZEUS \citep{Stone92a, Stone92b, Stone92c, Clarke96, Hayes06}. 
Such a study resolving sub-pc scales can be used to construct a subgrid model of AGN 
that could be used in large-scale galaxy simulations. 
\citet{KPN09} measured the energy, momentum, and mass feedback efficiencies due to radiation from AGN, 
and found that the values are smaller than what is often assumed ad hoc in cosmological simulations. 

The foundation for the present work is based on the tests of the ZEUS code, 
which were hydrodynamic simulations of spherically-symmetric Bondi accretion 
onto SMBHs \citep{PB03a, Proga07}.  See also \citet{ParkRicotti10} for an intermediate mass BH. 
Once a code has passed the test of 
successfully reproducing the simplest case of spherical Bondi accretion, it can be applied to related problems in various systems. 
Other studies have included additional physics to the problem; 
e.g., \citet{PB03a, PB03b} incorporated rotation and magnetic fields, 
while \citet{Moscibrodzka08, Moscibrodzka09} explored the effects of adiabatic index and gas temperature 
on hydro- and magneto-hydrodynamic accretion flows. 

Other authors have also used Bondi accretion to test various codes  
\citep[see][for a review]{Edgar04}. 
In a series of papers, \citet{Ruffert94} 
investigated the hydrodynamics of 3-D classical Bondi-Hoyle accretion using a grid-based code. 
Bondi-Hoyle-Lyttleton accretion have been studied in different systems, e.g., 
onto a protoplanetary disk using SPH \citep{Moeckel09}, 
for binary black hole mergers in gaseous environments using general-relativistic hydrodynamics \citep{Farris10}, 
in the early Universe before the formation of the first stars and galaxies \citep{Ricotti07}. 

We plan to follow the same path of first testing the spherical Bondi accretion scenario using SPH, before addressing the problem of AGN feedback. 
One of our long term goals is to confirm the results of \citet{KPN09} involving the feedback efficiencies. 


This paper is organised as follows. 
We describe our numerical method and simulation setup in \S\ref{sec-numerical}. 
We present and discuss the results in \S\ref{sec-results}. 
We summarise and conclude in \S\ref{sec-conclusion}.

\section{Numerical Method} 
\label{sec-numerical} 

\subsection{Revisiting the Bondi Problem} 
\label{sec-numerical-Bondi} 

The problem of spherically symmetric accretion onto a central mass was analysed in the seminal work by \citet{Bondi52}. 
It describes a central star at rest in a cloud of gas. 
The gas is at rest at infinity, where it is parametrised by uniform density $\rho_{\infty}$ and pressure $p_{\infty}$. 
The motion of the gas is steady and spherically symmetrical as it accretes onto the central star. 
The increase in mass of the star is ignored, so that the force field remains constant. 
The gas pressure $p$ and density $\rho$ are related by the 
polytropic equation of state, $p \propto \rho^\gamma$, 
with the adiabatic index satisfying $1 \leq \gamma \leq 5/3$. 

Two equations are solved for the gas motion to obtain the velocity $v$ and density as a function of radius $r$. 
First, the equation of mass continuity, 
\begin{equation} 
\label{eq-Continuity} 
\dot{M} = - 4 \pi r^2 \rho v = {\rm constant}.
\end{equation} 
where $\dot{M}$ is the mass accretion rate. Second, the Bernoulli's equation, which reduces to 
\begin{equation} 
\label{eq-Bernoulli} 
\frac{v^2}{2} + 
\left( \frac{\gamma}{\gamma - 1} \right) \frac{p_{\infty}}{\rho_{\infty}} \left[ \left( \frac{\rho}{\rho_{\infty}} \right)^{\gamma - 1} - 1 \right] 
= \frac{G M_{BH}}{r} ,
\end{equation} 
where the right-hand-side represents a Newtonian gravitational potential of the central star, 
which is a BH for our case. 
Several types of solution are possible (Figure~2 of \citealt{Bondi52}); 
the one relevant for astrophysical accretion is    
the so-called critical solution. 
In this solution, gas is subsonic in the outer parts, passes through a sonic point, 
and accretes onto the central object with a supersonic velocity. 
The mass accretion rate for such a motion is 
\begin{equation} 
\label{eq-MdotBondi} 
\dot{M}_{B} = 4 \pi \lambda_c \frac{ \left( G M_{BH} \right)^2 } { c_{s,\infty}^3 } \rho_{\infty}, 
\end{equation} 
with 
\begin{equation} 
\lambda_c = 
\left( \frac{1}{2} \right)^{ \frac{ \left( \gamma + 1 \right) } { 2 \left( \gamma - 1 \right) } }  
\left( \frac{5 - 3 \gamma}{4} \right)^{ \frac{ \left( 3 \gamma - 5 \right) } { 2 \left( \gamma - 1 \right) } } . 
\end{equation} 
Here $c_{s} = \sqrt{ \gamma k_B T / \left( \mu m_p \right) }$ is the sound speed in the gas of 
temperature $T$ and mean molecular weight $\mu$. 
Solving Eqs.~(\ref{eq-Continuity}) and (\ref{eq-Bernoulli}) gives the density and velocity of the Bondi solution, 
which we denote as $\rho_B (r)$ and $v_B (r)$. 

This analysis gives a characteristic length scale, the Bondi radius, 
\begin{equation} 
\label{eq-RBondi} 
R_B = \frac{ G M_{BH} } { c_{s,\infty}^2 } . 
\end{equation} 
The location of the sonic point can be expressed analytically as 
\begin{equation} 
\label{eq-Rsonic} 
R_s = \left( \frac{5 - 3 \gamma}{4} \right) R_B .  
\end{equation} 
An important timescale is the sound crossing time from a distance $R_B$ to the center (the Bondi time): 
\begin{equation} 
\label{eq-tBondi} 
t_B = \frac{R_B}{c_s} = \frac{ G M_{BH} } { c_{s,\infty}^3 }. 
\end{equation} 
The latter equality in Eq.~(\ref{eq-tBondi}) is for an isothermal case. 
These equations are for a purely Newtonian gravitational potential (Eq.~\ref{eq-Bernoulli}). 
As $\gamma \rightarrow 5/3$,     
the sonic radius asymptotically goes to zero ($R_s \rightarrow 0$), i.e., there is no relevant sonic point. 

However, for a problem of BH accretion the general-relativistic gravitational field differs from the Newtonian form at very small radii. 
The pseudo-Newtonian Paczynsky-Wiita potential 
(which we describe in \S\ref{sec-numerical-ModelSetup}, Eq.~\ref{eq-PsiPW}) can capture the relativistic effects well. 
For the Paczynsky-Wiita potential as well as for the fully general-relativistic problem \citep{Begelman78}, 
the Bondi flow with $\gamma = 5/3$ has a sonic point at roughly the geometrical mean 
of the Bondi radius and the Schwarzschild radius \citep[see also][]{PB03a}.

\subsection{Model Setup} 
\label{sec-numerical-ModelSetup} 

Our simulation setup consists of a spherical distribution of gas accreting onto a central SMBH with a mass of $M_{BH} = 10^8 M_{\odot}$. 
The inner and outer radii of our computational volume are chosen such that 
the Bondi and sonic radii lie well within our simulation domain. 
We choose the inner radius of $r_{\rm in} = 0.1$ pc, which is at least an order of magnitude smaller than the values of $R_s$ we explored. 
The outer radius is varied depending on the other model parameter values, 
and we explore a range of $r_{\rm out} = 5 - 200$ pc. 

We use the 3D Tree-PM Smoothed Particle Hydrodynamics code GADGET-3 \citep[originally described in][]{Springel05}. 
There are only gas particles in our simulations, 
because our outer radius goes only up to $5 - 200$ pc, and the dark matter density is much lower 
than the gas density in the central 10's - 100\,pc of a galaxy. 

The central SMBH is represented by the pseudo-Newtonian potential given by \citet{Paczynsky80}: 
\begin{equation} 
\label{eq-PsiPW} 
\Psi_{PW} = - \frac{ G M_{BH} } {r - R_g}, ~~~{\rm with}~~ R_g = \frac{ 2 G M_{BH} } {c^2}. 
\end{equation} 
Here $R_g$ is the gravitational radius of the BH. 
In our simulations, $R_g = 2.96 \times 10^{13}$ cm $ = 9.57 \times 10^{-6}$ pc; 
therefore the Paczynsky-Wiita potential is essentially Newtonian within our computational domain. 
This is represented in the GADGET code by a ``static potential'' approach, 
with the following acceleration added to each particle, 
\begin{equation} 
\vec{a}_{PW} = - \frac{ G M_{BH} } {(r - R_g)^2} \hat{r}. 
\end{equation} 
We also tested the effect of adding a galaxy bulge potential in our simulation, which is described in \S\ref{sec-results-Bondi-Inflow}. 
In our simulations we set the gravitational softening length of gas to values in the range $0.005 - 0.02$ pc. 
The minimum gas smoothing length is set to 0.1 of the softening lengths, which is $0.0005 - 0.002$ pc.

\subsection{Initial and Boundary Conditions} 
\label{sec-numerical-InitBound} 

We start with a spherical distribution of particles between $r_{\rm in}$ and $r_{\rm out}$, 
distributed according to the initial profiles of density $\rho_{\rm init} (r)$, velocity $v_{\rm init} (r)$, and temperature $T_{\rm init} (r)$. 
For most of our runs the initial profiles are taken from the Bondi solution, 
which is parametrised by the density $\rho_{\infty}$ and temperature $T_{\infty}$ at infinity. 
The initial conditions (ICs) are generated using an adiabatic index $\gamma_{\rm init}$, 
and the simulations are run with $\gamma_{\rm run}$. 
Most of our runs have $\gamma_{\rm init} = \gamma_{\rm run}$ (see Tables~\ref{Table-Bondi} and \ref{Table-HeatCool}). 
The values of different parameters we used, along with their justification are described in 
\S\ref{sec-results-Bondi} and \S\ref{sec-results-HeatCool}. 

Any particle going out of our computational domain ($r_{\rm in} < r < r_{\rm out}$)  
is considered to have escaped the boundary, and is removed from the simulation. 
Particles going inside $r_{\rm in}$ are being accreted into the inner boundary, 
and are counted in the mass inflow rate (\S\ref{sec-results-Bondi-Inflow}). 
Effectively, we simulate a static sink of radius $r_{\rm in}$, which absorbs the accreting particles. 
We tested some other outer boundary conditions that are discussed in \S\ref{sec-results-Bondi-Discussion}.

\subsection{Radiative Heating and Cooling} 
\label{sec-numerical-RadHC} 

Radiation from the central SMBH is considered to be in the form of a spherical 
X-ray emitting corona \citep[e.g.,][]{Proga07, POK08, KP09a}, which irradiates the accretion flow. 
The X-ray luminosity, $L_X$, is a fraction $f_{X}$ of the Eddington luminosity, $L_{\rm Edd}$: 
\begin{equation} 
L_X = f_{X} L_{\rm Edd},~~~~~L_{\rm Edd} = \frac{ 4 \pi c G m_p M_{BH} } {\sigma_e}, 
\end{equation} 
where $c$ is the speed of light, $G$ is the gravitational constant, $m_p$ is the proton mass, 
and $\sigma_e$ is the Thomson cross section for the electron. 
The local X-ray radiation flux at a distance $r$ from the central source is 
\begin{equation} 
F_X = \frac{L_X}{4 \pi r^2}. 
\end{equation} 
The heating-cooling function is parametrised in terms of the photoionization parameter, $\xi$, which is defined as 
\begin{equation} 
\label{eq-xi} 
\xi \equiv \frac{4 \pi F_X}{n} = \frac{L_X}{r^2 n}, 
\end{equation} 
where $n = \rho / (\mu m_p)$ is the local number density of gas. 
We use a hydrogen mass fraction of $0.76$ to estimate the mean molecular weight $\mu$. 

We include radiative processes in our simulations using the heating-cooling function from \citet{PSK00}. 
The equations are originally from \citet{Blondin94}, 
who presented approximate analytic formulae for the heating and cooling rates of an X-ray irradiated 
optically-thin gas illuminated by a $10$ keV bremsstrahlung spectrum. The net heating-cooling rate, ${\cal{L}}$, is given by 
\begin{equation} 
\label{eq-netL} 
\rho {\cal{L}} = n^2 \left( G_{\rm Compton} + G_{X} - L_{b,l} \right) \quad [{\rm erg~cm^{-3}~s^{-1}}], 
\end{equation} 
where each of the components are formulated below. 
The rate of Compton heating and cooling, 
\begin{equation} 
\label{eq-GCompton} 
G_{\rm Compton} = 8.9 \times 10^{-36} \xi \left( T_X - 4T \right) \quad [{\rm erg\,cm^3\,s^{-1}}]. 
\end{equation} 
The net rate of X-ray photoionization heating and recombination cooling, 
\begin{equation} 
G_{X} = 1.5 \times 10^{-21} \xi^{1/4} T^{-1/2} \left( 1 - \frac{T}{T_X} \right) \quad [{\rm erg~cm^3~s^{-1}}]. 
\end{equation} 
The rate of bremsstrahlung and line cooling, 
\begin{eqnarray} 
\label{eq-Lbl} 
L_{b,l} & = & 3.3 \times 10^{-27} T^{1/2}  \nonumber  \\ 
&& + \left[ 1.7 \times 10^{-18} \exp \left( - 1.3 \times 10^{5} / T \right) \xi^{-1} T^{-1/2} \right.  \nonumber  \\ 
&& \left. ~~~~ + 10^{-24} \right] \delta \quad\quad [{\rm erg~cm^3~s^{-1}}]. 
\end{eqnarray} 
We adopt the optically thin version of line cooling in Eq.~(\ref{eq-Lbl}) by setting $\delta = 1$. 
In the above, $T_X$ is the characteristic temperature of the bremsstrahlung radiation. 
We use $T_X = 1.16 \times 10^{8}$ K, corresponding to \citet{Blondin94}'s value of $10$ keV. 

The heating-cooling rate is a function of $\xi$, $T_X$ and $T$. 
In the code, ${\cal{L}}$ is computed for each active particle, 
and added to the specific internal energy (entropy in GADGET-3) equation of each particle 
using a semi-implicit method. 
In the entropy equation, the non-radiative terms are integrated in an explicit fashion using the simulation timestep, 
and then the radiative term is integrated using an implicit method. 
This integration methodology is the same as that of 
radiative cooling and photoionization heating in a cosmological context in the GADGET code \citep[e.g.,][]{Katz96}.

\section{Results and Discussion}
\label{sec-results} 

\subsection{Reproducing Bondi Accretion} 
\label{sec-results-Bondi} 

First, we perform a series of simulations of Bondi accretion (i.e., there is no radiative heating and cooling) 
as listed in Table~\ref{Table-Bondi}. 
We use $T_{\infty} = 10^7$ K and $\rho_{\infty} =  10^{-19}$ g/cm$^3$, 
which are typical values at $10$'s of pc away from SMBH used in AGN accretion simulations \citep[see e.g.,][and references therein]{KP09b}. 
Since $R_s \rightarrow 0$ as $\gamma \rightarrow 5/3$ (Eq.~\ref{eq-Rsonic} in \S\ref{sec-numerical-Bondi}), 
we use $\gamma_{\rm init} = 1.01$ in order to have the Bondi and sonic radius well between $r_{\rm in}$ and $r_{\rm out}$. 
Therefore the equation of state is almost isothermal, 
and the simulations are run with the same value of $\gamma_{\rm run} =  1.01$. 
For these parameters, the Bondi radius is at $R_B = 3.0$ pc, 
the theoretical value of the sonic point is $R_s = 1.5$ pc, 
and the Bondi time is $t_B = 7.9 \times 10^3$ yr. 

All the runs in Table~\ref{Table-Bondi} have $T_{\rm init} (r) = T_{\infty}$. 
In the Bondi IC runs, the initial condition is generated from the Bondi solution, 
i.e., the initial particles follow $\rho_{\rm init} (r) = \rho_B (r)$ and $v_{\rm init} (r) = v_B (r)$. 
In the uniform IC runs, we start with a constant initial density of $\rho_{\infty}$ and $v_{\rm init} = 0$.

\subsubsection{Particle Properties} 
\label{sec-results-Bondi-Prop} 

Figure~\ref{fig-BondiScatterVsR} is a scatter plot showing the 
properties of particles vs.~radius in a representative Bondi accretion simulation, Run 7, which has $r_{\rm out} = 20$\,pc and the Bondi IC. 
The radial component ($v_r$) of the velocity $v$ and the density profile 
follows the Bondi solution quite well, except near the inner and outer radii. 
A negative value of $v_r$ represents inflowing mass, whereas positive $v_r$ denotes an outflow. 
We do not show the non-radial velocity components (i.e., $v_{\theta}$ and $v_{\phi}$) 
because they are typically $100 - 1000$ times smaller than $v_r$. 
The temperature profile is almost isothermal at $10^7$ K (= $T_{\infty}$), 
which is expected since we used  $\gamma_{\rm run} =  1.01$. 
Examining the Mach number profile, we see that the gas is subsonic near $r_{\rm out}$, 
passes through a sonic point, and approaches $r_{\rm in}$ with supersonic velocity (Mach $= 6$). 
The location of the sonic point (where the gas crosses Mach $= 1$) in the simulation is $\sim 1.5$ pc, 
consistent with the theoretical value of the Bondi solution ($R_s$, \S\ref{sec-results-Bondi}). 
The smoothing length ($h_{sml}$) of particles near  $r_{\rm in}$ is $\sim 0.12$ pc. 
It is much larger than the minimum gas smoothing length, which is set to $0.0005$\,pc in this run.  
This finite numerical resolution is partly responsible for the deviations of the profiles from the ideal Bondi solution at small radii. 

The slight decrease of the density from $\rho_B (r)$ and the corresponding increase in smoothing length at $r < 0.13$\,pc 
is an artifact of our inner boundary condition. 
There are no particles inside $r_{\rm in}$, as they accrete onto the sink. 
Therefore the smoothing spheres of the particles just outside of $r_{\rm in}$ 
overlap with the sink, causing them to miss some neighbors, and the density is slightly underestimated. 
\citet{Bate95} corrected the boundary conditions near the sink particle in SPH as a solution to this problem. 
Since this effect appears only at $r < 0.13$\,pc in our simulations, 
we do not adopt any special boundary conditions for the sink. 

There is a group of particles flowing out of the computational volume at $r_{\rm out}$, 
because of finite pressure gradient there. 
According to our outer boundary condition, 
there are no particles just outside $r_{\rm out}$, 
and particles just inside $r_{\rm out}$ feel an outward pressure gradient and are pushed out. 
This explains the following features at $r > 10$ pc: 
$v_r > 0$, increasing Mach, decreasing $\rho$ and increasing $h_{sml}$. 
We see that this spurious, unwanted outflow depends significantly on the gas temperature. 
Most of our simulations with a higher temperature $T_{\infty} = 10^7$ K 
(all in Table~\ref{Table-Bondi}, few in Table~\ref{Table-HeatCool}) show this outflow prominently,  
while the simulations with lower $T_{\infty} = 10^5$\,K have much less outflow. 
This is because a lower temperature at $r_{\rm out}$ leads to a lower gas pressure, 
which reduces the outward pressure gradient acting on the particles, and weakens the mass outflow at $r_{\rm out}$. 

The radial properties of all the runs in Table~\ref{Table-Bondi} look similar to Figure~\ref{fig-BondiScatterVsR}. 
The small scatter of the quantities vs.~radius shows that the accretion flow is spherically symmetric, 
especially in the inflowing regions.

A cross-section slice of gas density is presented in Figure~\ref{fig-Bondi-Splash} for Run 4, 
which has $r_{\rm out} = 5$\,pc and starts with the Bondi IC. 
It shows the $X - Y$ plane, with the cross-section through $Z = 0$. 
The velocity vectors of the gas are overplotted as arrows. 
This cross-section image shows that the flow is indeed nearly spherically symmetric. 
The velocity vectors are symmetrically pointing inward in the inner volume, where there is net inflow of particles.

\subsubsection{Mass Flux Evolution at the Inner Boundary}  
\label{sec-results-Bondi-Inflow} 

The mass flux of accreting gas at a given time as a function of radius can be expressed as 
\begin{equation} 
\label{eq-MassFlux1} 
\dot{M} (r) = \oint_{S} \rho ~ v ~ \cdot ~ da = r^2 \oint_{4 \pi} \rho ~ v_r ~ d\Omega. 
\end{equation} 
In particular, we compute the mass inflow rate at the inner boundary, $\dot{M}_{{\rm in}, r_{\rm in}}$, 
and consider it as the figure of merit to determine 
how well a simulation can reproduce the Bondi accretion. 
We sum up the mass of particles accreted inside $r_{\rm in}$ within a certain time interval ($\Delta t$), 
then compute the inflow rate as: 
\begin{equation} 
\dot{M}_{{\rm in}, r_{\rm in}} (t) = \frac{1}{\Delta t} \sum_{(r_k < r_{\rm in})} M_{k}. 
\end{equation} 
The GADGET code uses individual timesteps for particles, so not all particles are active at once. 
Using the simulation timestep as $\Delta t$ produced very spiky $\dot{M}_{{\rm in}, r_{\rm in}}$ over time, 
therefore we sum over the masses of particles 
accreted during $128$ or $256$ timesteps for the results presented in the following. 
If a simulation can track Bondi accretion perfectly, it would have $\dot{M}_{{\rm in}, r_{\rm in}} = \dot{M}_{B}$ (Eq.~\ref{eq-MdotBondi}). 
{\it We check how closely and for how long this relation is satisfied for our runs.} 

The mass inflow rate as a function of time is plotted in Figs.~\ref{fig-Bondi-Mdot_Rout_N} 
and \ref{fig-Bondi-Mdot_IC_Bulge} for the runs in Table~\ref{Table-Bondi}. 
The horizontal straight line in each panel indicates the corresponding Bondi accretion rate 
$\dot{M}_{B} = 4.6 \times 10^{27}$ g/s. 

Figure~\ref{fig-Bondi-Mdot_Rout_N} shows the runs with the Bondi IC 
and a uniform IC (constant density and velocity), 
the first eight runs in Table~\ref{Table-Bondi}. 
We see that the runs starting with the Bondi IC reach steady-state quickly, 
and can reproduce the Bondi accretion rate for some time.   
This time range increases with increasing outer radius of the computational volume. 
Runs with $r_{\rm out} = 5, 10, 20, 50$\,pc follow the Bondi rate for time durations of 
$t \approx (0.6, 1.6, 4, 8) \times 10^{4}$ yrs (i.e., $0.76, 2.0, 5.1, 10.1 ~ t_B$), respectively. 
The run with $r_{\rm out} = 10$\,pc is tested with 2 numerical resolutions, using particle numbers of $N = 64^3$ and $128^3$. 
There is no significant difference in the $\dot{M}_{{\rm in}, r_{\rm in}}$ vs.~time, 
except that the $N = 64^3$ run has a larger scatter of the mass inflow rate because of a lower mass resolution. 
The uniform IC runs start with a low $\dot{M}_{{\rm in}, r_{\rm in}}$ at early times 
(lower than the corresponding Bondi IC run by more than an order of magnitude), have an increase, 
but cannot catch up with the Bondi rate in the runs with $r_{\rm out} = 5$ or $10$ pc. 
With a larger outer boundary, $r_{\rm out} = 50$ pc, 
the uniform IC run eventually starts to follow the Bondi IC run after $t = 3 \times 10^{4}$ yrs = $3.8 ~ t_B$. 

The top panel of Figure~\ref{fig-Bondi-Mdot_IC_Bulge} 
gives the results of starting with the gas at rest ($v_{\rm init} = 0$) and having different initial gas density profiles: 
$\rho_B (r)$, and a uniform density ($\rho_{\infty}$). 
The mass inflow rates for the Bondi and uniform cases increase with time, 
reaches a maximum value (almost steady-state) of $\dot{M}_{B}$, then falls off after $4 \times 10^{4}$ yrs. 

The main reason for the drop in $\dot{M}_{{\rm in}, r_{\rm in}}$ after few $\times 10^{4}$ yrs 
in all of our runs is that the particles are lost from the simulation volume, 
as described in \S\ref{sec-results-Bondi-Prop}, 
and there are not enough particles between $r_{\rm in}$ and $r_{\rm out}$ to correctly represent the Bondi problem. 
Our simulations start with $N$ particles (third columns in the tables) as the IC. 
Then according to our boundary conditions (\S\ref{sec-numerical-InitBound}), 
particles accreted into $r_{\rm in}$ or moving outside $r_{\rm out}$ are removed from the simulation. 
This causes the particle number (and hence the total mass) between $r_{\rm in}$ and $r_{\rm out}$ to continuously decrease with time. 
Despite this mass loss, several of our runs reached a steady-state 
for durations of a few $\times 10^{4}$ yrs, 
i.e., a few Bondi times (Eq.~\ref{eq-tBondi}), which is enough for the Bondi problem we are studying. 

This loss of particles occurs mainly out of $r_{\rm out}$ for the runs in Table~\ref{Table-Bondi}. 
We already discussed this outflow in \S\ref{sec-results-Bondi-Prop}, 
that it is due to the high temperature $T_{\infty} = 10^7$ K. 
As Table~\ref{Table-Bondi} shows, 
the cumulative mass fraction accreted into $r_{\rm in}$ is   
between $2 - 35 \%$ of the initial mass, by the end time of the simulation. 
The majority of mass ($55 - 90\%$) move out of $r_{\rm out}$ by the end of simulation. 
In order to remedy this problem, we tried other approaches of handling the outer boundary condition, 
which we discuss in \S\ref{sec-results-Bondi-Discussion}. 

The above results demonstrate that the simulations starting with the Bondi IC 
reach steady-state sooner than those starting with a uniform IC. 
For our next set of runs in Table~\ref{Table-HeatCool}, 
we start with the Bondi IC in order to reach the desired steady-state solution faster.

\begin{twocolumn} 

\subsubsection{Hernquist Density \& Bulge Potential} 

In order to explore the dependence of our results on the initial gas density profile, 
in Run 11 we start with a different density: the \citet{Hernquist90} profile $\rho_H (r)$ and $v_{\rm init} = 0$. 
The density profile has a radial dependence of $\rho_H (r) \propto r^{-1} (r + a_{\rm bulge})^{-3}$, 
which is normalized to contain the same value of total mass within our computational volume as with the Bondi density profile. 
The corresponding mass inflow rate is shown in the top panel of Figure~\ref{fig-Bondi-Mdot_IC_Bulge}, 
starting with $\dot{M}_{{\rm in}, r_{\rm in}} = 1.4 \dot{M}_{B}$ and then a decrease with time, never reaching a steady-state. 
This is because $\rho_H (r)$ is steeper than $\rho_B (r)$, 
therefore more mass is concentrated in the inner parts of the simulation volume, causing faster accretion.

The bottom panel of Figure~\ref{fig-Bondi-Mdot_IC_Bulge} 
shows the effect of adding a galaxy bulge potential to the Bondi accretion simulation setup. 
We consider a bulge potential for the \citet{Hernquist90} profile:  
\begin{equation} 
\label{eq-PsiH} 
\Psi_{H} = - \frac{ G M_{\rm bulge} } { r + a_{\rm bulge} }, 
\end{equation} 
where $M_{\rm bulge} = 3.4 \times 10^{10} M_{\odot}$ is the mass and $a_{\rm bulge} = 700$\,pc is the scale length 
of a Milky-Way type galaxy bulge \citep{Johnston96}. 
In Run 12, we start with the Bondi IC, and add $\Psi_{H}$ to the simulation    
using a static-potential approach (\S\ref{sec-numerical-ModelSetup}). 
Since $a_{\rm bulge} = 700$\,pc is significantly larger than the outer boundary $r_{\rm out} = 20$ pc, 
we are effectively adding a constant gravitational potential $\Psi_{H} \simeq - GM_{\rm bulge} / a_{\rm bulge}$ to the simulation. 
We see that this increases the mass inflow rate with respect to the Bondi rate, 
which is expected because the particles fall inward and are accreted more readily. 
The value of $\dot{M}_{{\rm in}, r_{\rm in}}$ rises with time, but does not attain a proper steady-state, 
and reaches a maximum value of $1.6~\dot{M}_{B}$ before falling off.

\subsubsection{Mass Flux as a Function of Radius} 
\label{sec-results-Bondi-MassFlux} 

We also measure the mass flux of gas versus radius in order to examine 
how closely the flow follows the Bondi accretion rate. 
In our SPH simulations velocities and densities are computed for discrete particles, therefore we determine the mass flux as   
\begin{equation} 
\label{eq-MassFlux2} 
\dot{M} (r) = \frac{ 4 \pi r^2 } {N_{\rm sph}} \sum_{(r_k > r, ~ r_k-h_{sml, k} < r)} \rho_{k} ~ v_{r, k} . 
\end{equation} 
The summation is over $N_{\rm sph}$ particles touching the spherical surface area 
at a radius $r$ through their smoothing lengths. 
The net mass flux ($\dot{M}_{\rm net}$) is obtained above if all $v_r$ are included in Eq.~(\ref{eq-MassFlux2}). 
The inflow mass flux ($\dot{M}_{\rm in}$) can be calculated by only counting particles with $v_r < 0$, 
and the outflow mass flux ($\dot{M}_{\rm out}$) by counting particles with $v_r > 0$. 

The resulting mass flux is plotted in Figure~\ref{fig-Bondi-MassFlux} as a function of radius 
for the four runs  in Table~\ref{Table-Bondi} with Bondi IC. 
In the inner regions ($0.1 < r <$ a few pc), all the mass is inflowing at the Bondi rate, 
i.e., $\dot{M}_{\rm net} = \dot{M}_{\rm in} = \dot{M}_{B}$, and $\dot{M}_{\rm out} = 0$. 
Near the outer boundary, 
the inflow rate reduces and outflow increases toward $r_{\rm out}$. 
Here the net mass flux is dominated by this outflow, which is greater than the Bondi rate for $r_{\rm out} \geq 10$ pc. 
We find that the transition radii ($r_{tr}$) from net inflow to net outflow varies with time for a given run. 
It starts at the outer boundary, and comes inward with the progress of time, as more and more particles flow out. 

In Run 7, all the gas between $r = 0.1$ and $8$\,pc is inflowing at time $0.016$ Myr. 
Beyond $8$\,pc there is an outflow, but the inflow still dominates the mass flux until $r_{tr} = 12$ pc. 
After $r_{tr}$, the net mass flux is dominated by the outflow of the gas. 
The results for different outer radii are qualitatively similar to each other.

\subsubsection{Different Outer Boundary Conditions} 
\label{sec-results-Bondi-Discussion}

In an attempt to remedy the above mentioned outflow problem,    
we tried some alternate approaches of our outer boundary condition,     
two of which were successful in reducing the outflow, as discussed below.       
With a simulation setup similar to ours,   
\citet{Cuadra06} modeled the accretion of stellar winds on to Sgr A* using the GADGET code.   
They also find an outflow,    
at least $99\%$ of their `wind' (gas) particles escaping from the inner volume into the greater Galaxy.  

The numerical representation of the initial conditions used in our code does not guarantee 
that the Bernoulli's function (Eq.~\ref{eq-Bernoulli}, LHS - RHS) is {\it exactly} zero everywhere.  
Therefore we check whether there are any unbound particles in the initial conditions, and if so what are their effects. 
We found that the fraction of initially unbound particles depends on the initial scatter of the radial density profile. 
But these unbound particles have no significant effect on the total outflowing mass fraction. 
We performed a few more tests starting with an initial condition with $< 5 \%$ scatter (at a fixed-$r$) 
in the Bondi density profile, and one with a perfectly uniform IC. 
These runs also have the outflow, with the same mass fraction flowing out of $r_{\rm out}$ as the larger initial-scatter runs. 
This verifies that the outflow is being caused primarily by our outer boundary condition, 
i.e., the outward pressure gradient on particles near $r_{\rm out}$ because of missing neighbors. 


We note that others also found 
that setting up the desired boundary conditions can be very challenging in SPH \citep[e.g.,][]{Herant94}.   
The main reason for this is that the quantities, 
of which conditions are to be set, are determined by smoothing over neighboring particles, 
which could be located anywhere with respect to the boundaries. 
For our case it is the difficulty of implementing the Bondi assumption of an infinite and spherical reservoir of gas at infinity. 
In an Eulerian code like ZEUS the gas density at the outer boundary can be easily held fixed at a constant value,   
therefore an outer radius as small as $1.2 R_B$ is enough to adequately reproduce Bondi accretion \citep[see][]{PB03a}. 


A similar outflow issue is dealt with in MHD SPH simulations of molecular cloud core collapse 
by placing the cloud within a uniform, low-density box of surrounding material in pressure equilibrium, 
where the outer box size is twice the cloud radius in each direction \citep{Price07}. 

We are successful in reducing the outflow to a negligible fraction, still maintaining spherical-symmetry of the accreting gas,   
by altering some property of every particle which has a net outward radial velocity ($v_r > 0$)   
in a wide outer shell: between $0.4 r_{\rm out}$ and $r_{\rm out}$, in a run with $r_{\rm out} = 20$ pc.    
Below we describe the runs with two of the stringent conditions which can successfully reduce the outflow. 
Runs 7a and 7b have everything similar to Run 7, but with two different implementations of the outer boundary condition.  

(1) In Run 7a, whenever a particle between $[0.4 r_{\rm out} - r_{\rm out}]$ has $v_r > 0$,    
we change its velocity to the Bondi velocity, i.e., we set $v_r = v_B$, 
keeping everything else ($r, \theta, \phi,$ etc.) same.  
Since $v_B$ is negative, the outward flowing particle is forced to move inward, on imposing this condition.  
As listed in Table~1, such a boundary condition can reduce the outflowing mass significantly: 
65\% mass is accreted into $r_{\rm in}$ and 4\% flows out of $r_{\rm out}$, within $100 t_B$. 

(2) Run 7b has a layer of particles at $r_{\rm out}$ which do not feel the net outward pressure gradient 
because of missing neighbors just outside $r_{\rm out}$. 
We alter the $(\nabla P / \rho)_i$ term directly in the particle acceleration $dv_i / dt$ equation. 
We set a zero pressure gradient, $\langle \nabla P / \rho \rangle_i = 0$, for all particles between $[0.9 r_{\rm out} - r_{\rm out}]$  
and for those which have $v_r > 0$ in between $[0.4 r_{\rm out} - r_{\rm out}]$. 
This run has a further reduction in the outflowing mass: 93\% accretes in, and only 1\% moves out, within $125 t_B$. 

Figure~\ref{fig-Bondi-Mdot_OuterBC} shows the mass inflow rate as a function of time of Runs 7a and 7b. 
We see some initial unsteadiness in both the runs manifesting as spikes in $\dot{M}_{{\rm in}, r_{\rm in}}$. 
After $0.08$ Myr, the mass inflow becomes nearly-steady ($\simeq \dot{M}_{B}$) for $\sim 12 t_B$ in Run 7a and $\sim 25 t_B$ in Run 7b. 
$\dot{M}_{{\rm in}, r_{\rm in}}$ decreases after that; 
now this reduction is because a dominant mass fraction is being accreted into $r_{\rm in}$.     

Such an outer boundary condition (as in Run 7b) can be adopted with any of the runs presented in this paper    
to obtain a longer duration for which steady-state Bondi accretion continues onto $r_{\rm in}$.   
All the gas properties remain the same as presented here, for that longer duration.

Table~\ref{Table-Rout-Time} shows some relevant time durations for the Bondi IC simulation runs, 
with increasing $r_{\rm out}$. 
The ratio of the total initial mass to the Bondi accretion rate, $M_{\rm tot,IC} / \dot{M}_{B}$, 
gives an upper limit to the time duration for which the simulation gives reliable results. 
$M_{\rm tot,IC} / \dot{M}_{B}$ is the time within which all mass would be accreted in, assuming accretion at the Bondi rate. 
In practice the significant time duration is rather a fraction of $M_{\rm tot,IC} / \dot{M}_{B}$, 
as we see in our runs, 
because there has to be some particles remaining in the simulation to reproduce the Bondi problem.

Summarising the results, our simulations can reproduce adiabatic Bondi accretion within a limited time duration, 
ranging between one to ten Bondi time, which is long enough to investigate the Bondi problem. 
The radial profiles of gas density, velocity, temperature, and Mach number 
match the Bondi solutions quite well, except near the inner and outer radii.

\subsection{Spherical Accretion with Radiative Heating and Cooling} 
\label{sec-results-HeatCool} 

In the next series of simulations we include radiative heating and cooling. 
These runs are relevant to the studies of feedback due to radiative heating in the context of SMBH accretion. 
Table~\ref{Table-HeatCool} gives a list of the runs, where we follow the prescriptions in \S\ref{sec-numerical-RadHC} 
and our spherical accretion simulation setup. 
All these runs are done with $\gamma_{\rm run} = 5/3$. 
Since $R_s = 0$\,pc for $\gamma_{\rm run} = 5/3$, 
the initial condition is generated using the Bondi solution with $\gamma_{\rm init} = 1.4$ 
in order to have the sonic point within current computational volume. 
We varied the X-ray luminosity in the range of $L_X / L_{\rm Edd} = 5 \times 10^{-5} - 0.5$. 
We also vary several other simulation parameters in our heating-cooling runs 
($r_{\rm out}$, $N$, $\gamma_{\rm init}$, $T_{\infty}$, $\rho_{\infty}$, $T_{\rm init}$), 
as discussed in the subsequent sections. 

The radiative equilibrium condition can be obtained by solving for ${\cal{L}} = 0$ in Eq.~(\ref{eq-netL}). 
This gives a relation between the equilibrium temperature, $T_{\rm rad}$, and the photoionization parameter. 
The Compton equilibrium temperature is 
$T_C = T_X / 4 = 3 \times 10^7$ K, using $T_X$ from \S\ref{sec-numerical-RadHC}. 
This is comparable to the equilibrium Compton temperature of $2 \times 10^7$ K found in \citet{Sazonov04}. 
At high photoionization, $\xi > 10^5$, Compton processes dominate, resulting in a constant temperature $T_{\rm rad} = T_C$ for all $\xi$. 
At $\xi < 10^5$, other radiative cooling components start dominating and $T_{\rm rad}$ decreases below $T_C$. 


\subsubsection{A Representative Case} 
\label{sec-results-HC-CompareBondi} 

Run 23 is a representative simulation with radiative processes included, and 
the particle properties as a function of radius is plotted in Figure~\ref{fig-HC-ScatterVsR}. 
The velocity and density profiles are qualitatively similar to that in the Bondi accretion Run 7 in Figure~\ref{fig-BondiScatterVsR}; 
the gas fall inward with a larger velocity at smaller radii and its density rises. 

The temperature profile in Figure~\ref{fig-HC-ScatterVsR} shows signatures of heating and cooling operating at different radii, 
in contrast to the isothermal temperature profile of the Bondi problem. 
Within $r \sim 0.7 - 180$ pc, temperature is between $T \sim 10^4 - 10^5$ K and slowly increases inward. 
There is a sharp increase of temperature toward $r_{\rm in}$, 
and a slight heating near $r_{\rm out}$. 
The sonic point is close to the outer boundary at $r_{\rm out} = 200$ pc, and the gas in most of the volume is supersonic. 
The Mach number increases inward up to $\sim 30$ at $r \sim 0.7$ pc, but decreases at smaller $r$,  
as opposed to the monotonic increase in the Bondi accretion run. 
This is because of the steep heating near $r_{\rm in}$ in Run 23, 
which causes a greater increase in the sound speed than the increase in the dynamical gas velocity, resulting in a decrease of the ratio $|v| / c_s$. 

Supersonic gas (with $\gamma_{\rm run} = 5/3$) should only feel the gravity of the central SMBH effectively in this run, 
because the gas pressure would be negligible compared to the central gravity; 
therefore the gas should essentially be in the state of a free-fall.   
We compute the theoretical free-fall scaling of some of the gas properties, 
and compare those with the particle properties in our numerical simulations. 
The free-fall velocity for radial infall (in a purely Newtonian $1/r$ potential) is 
\begin{equation} 
\label{eq-Vff} 
\vec{v}_{ff} = -\sqrt{ \frac{ 2 G M_{BH} } {r} } \hat{r}. 
\end{equation} 
Then taking the equation of continuity, assuming steady-state, only-radial dependence of density, 
and using the above $\vec{v}_{ff}$, we obtain the free-fall density scaling, 
\begin{equation} 
\label{eq-Rho_ff} 
\frac{ \rho_{ff} } {\rho_0} = \left( \frac{r} {r_0} \right)^{-1.5}. 
\end{equation} 
Here, $\rho_0$ and $r_0$ are scaling constants, and the density is equal to $\rho_0$ at $r = r_0$. 
Using $\vec{v}_{ff}$ and $\rho_{ff}$ we obtain a free-fall adiabatic radial profile for the temperature 
by solving the equation of internal energy with only the adiabatic term and $\gamma = 5/3$, 
\begin{equation} 
\label{eq-T_ffa} 
T_{ff, a} \propto {r} ^{-1}. 
\end{equation} 
We also solve the general internal energy equation including both adiabatic and radiative terms, 
using $\vec{v}_{ff}$ and $\rho_{ff}$, and obtain the temperature solution $T_{ff,ar}$. 
The equations and details of the method are given in Appendix~\ref{sec-appenix}. 
Mach$_{ff,ar}$ is obtained by calculating the sound speed from $T_{ff,ar}$, 
and taking its ratio with the free-fall velocity. 

The free-fall scalings $v_{ff}$, $\rho_{ff}$, $T_{ff,ar}$ and Mach$_{ff,ar}$ are overplotted as the 
dashed curves along with the particle properties in Figure~\ref{fig-HC-ScatterVsR}. 
The values of the scaling parameters are manually set to $\rho_0 = 3 \times 10^{-23}$ g/cm$^3$ and $r_0 = 65$ pc, 
in order to get a good agreement. 
We see that the particle properties in our simulation match the theoretical free-fall solutions 
well in the intermediate radii, as predicted for supersonic gas. 
There is a deviation near the inner radius, where the gas velocity in our run is lower than $v_{ff}$,   
for which numerical resolution is partly responsible. 
Particles are heated up more than $T_{ff,ar}$ at $r < 0.7$ pc, 
and have a steep rise in temperature reaching $T = {\rm few} \times 10^7$ K at $r_{\rm in}$, 
compared with the expected value of ${\rm few} \times 10^5$ K. 
This is reflected as a decrease in Mach number at $r < 0.7$ pc, 
that drops to $\sim 2$ at $r_{\rm in}$, which is much smaller compared to the expected value of $30$. 
The region near the outer radius also shows a deviation from the free-fall scalings. 
This is caused by particles flowing outward at $r_{\rm out}$. 

In part the heating occurring near the inner boundary is caused by adiabatic compression of the gas flowing inward into smaller volumes. 
The free-fall temperature relation expected from adiabatic processes only ($T_{ff, a}$) 
is plotted as the dotted curve in Figure~\ref{fig-HC-ScatterVsR}. 
Comparing the temperature between the simulation and two predicted radial profiles, we see that 
the slope of $T$ vs. $r$ of the simulation particles near the inner radius is larger than both the predictions $T_{ff,ar}$ and $T_{ff, a}$. 
Therefore this inner heating cannot be fully explained by adiabatic and/or radiative processes. 
On further investigation (see below for more details) 
we find that this extra heating of the gas is caused by the artificial viscosity (AV) in the GADGET code. 
In Run 23, the AV heating dominates at $r < 0.7$ pc. 
The extra thermal energy comes at the expense of kinetic energy. 
The velocity of the particles is smaller than $v_{ff}$ at the same $r$, as the top-left panel of Figure~\ref{fig-HC-ScatterVsR} shows. 

The AV heating occurs within a radius much smaller than the sonic or critical radius, 
which is close to the outer boundary $r_{\rm out} = 200$ pc. 
Therefore the gas being heated up by AV is already supersonic, and remains so as it falls into $r_{\rm in}$. 
The Mach number reduces, however the gas does not become subsonic near $r_{\rm in}$, 
as the bottom-right panel in Figure~\ref{fig-HC-ScatterVsR} shows. 
So the gas temperature is affected by AV, but the central mass inflow rate is not affected because the gas is still supersonic. 

We perform 1-D simulation with the ZEUS code, using the same parameters as Run 23, for comparing with the GADGET run. 
The particles properties from ZEUS are overplotted in Figure~\ref{fig-HC-ScatterVsR} 
as the solid curve in the first three panels. 
We see that the ZEUS results follow the free-fall scalings well up to the inner radius. 
$v_{\rm ZEUS}$ is essentially indistinguishable from $v_{ff}$. 
$\rho_{\rm ZEUS}$ is at a constant offset from $\rho_{ff}$ because of different normalization. 
$T_{\rm ZEUS}$ goes to somewhat higher value than $T_{ff,ar}$, reaching $T_{\rm ZEUS} \sim 10^6$ K at $r_{\rm in}$; 
however not as high as the temperature in the GADGET run ($T = {\rm few} \times 10^7$). 
The difference between GADGET and ZEUS results of the temperature and velocity near $r_{\rm in}$ 
further confirms that the inner heating is a numerical artifact. 
It is due to AV, which is only present in the GADGET run.

The mass inflow rate at the inner radius of Run 23 is plotted in Figure~\ref{fig-HC-Mdot_Rout200pc} (bottom-right panel). 
$\dot{M}_{{\rm in}, r_{\rm in}}$ increases with time reaching a value $4~\dot{M}_{B}$ at $t = 2$ Myr, and continues to rise after that. 
It takes a longer time to reach a steady-state, compared to the Bondi accretion runs.  
However, Run 23 reaches almost steady-state at later times.

\subsubsection{Dependence on Model Parameters}    
\label{sec-results-HC-Inflow} 

We explored the effects of varying some model parameters:  
$r_{\rm out}$, $N$, $\gamma_{\rm init}$, $T_{\infty}$, $\rho_{\infty}$, and $T_{\rm init}$, 
in our spherical accretion simulations with radiative heating and cooling. 
The mass inflow rate vs.~time for the first four runs in Table~\ref{Table-HeatCool} with $T_{\rm init} = T_{\infty} = 10^7$ K 
is shown in Figure~\ref{fig-HC-Mdot_Rout_LX}. 
Runs 13 and 14 are shown in the top panel, for which $\rho_{\infty} = 10^{-21}$ g/cm$^3$ 
and the Bondi accretion rate is $\dot{M}_{B} (\gamma = 5/3, \rho_{\infty}, T_{\infty}) = 4.9 \times 10^{24}$ g/s. 
This panel gives results with different outer radii, 
and we find $\dot{M}_{{\rm in}, r_{\rm in}}$ ($r_{\rm out} = 50$\,pc) $< \dot{M}_{{\rm in}, r_{\rm in}}$ ($r_{\rm out}=20$\,pc) 
up to $t = 3 \times 10^4$ yrs, the time until which the $20$\,pc run has enough particles within the simulation volume to track the Bondi problem. 

The evolution of $\dot{M}_{{\rm in}, r_{\rm in}}$ for these four runs using $T_{\rm init} = 10^7$ K are qualitatively similar. 
The runs start with a high inflow rate of $1.4 \dot{M}_{B}$ (Run 14) and $2 \dot{M}_{B}$ (Runs 13, 15, 16), 
which is the maximum value they reach, and have a decline in $\dot{M}_{{\rm in}, r_{\rm in}}$ with time.   
Runs 13 and 14 never reach a steady-state. 
These runs have a very large mass fraction moving out of the computational volume, 
the largest among all the runs we have done. 
As given in Table~\ref{Table-HeatCool}, $98 - 99 \%$ of the total initial mass moves out of $r_{\rm out}$ by the end of the simulation, 
while only $0.1 - 0.5 \%$ of the total mass is accreted into $r_{\rm in}$. 
This is because the high initial temperature $T_{\rm init} = 10^7$ K heats up gas, which tends to expand and increase the outflow at $r_{\rm out}$. 
The mass outflow at the outer boundary is almost solely controlled by the gas temperature, 
because changing other parameters has almost no effect on it (\S\ref{sec-results-HC-Lx}). 

The mass inflow rate as a function of time for seven runs in Table~\ref{Table-HeatCool}, 
which has $T_{\rm init} = T_{\rm rad}$ (the radiative equilibrium temperature), is shown in Figure~\ref{fig-HC-Mdot_Rout200pc}. 
The time dependence of $\dot{M}_{{\rm in}, r_{\rm in}}$ of these runs is similar to each other;  
it increases with time, reaches a maximum value, then falls off after a certain time. 
Runs with $r_{\rm out} = 20$\,pc have an increase in $\dot{M}_{{\rm in}, r_{\rm in}}$ up to $\sim 10^5$ yrs, and 
for runs with $r_{\rm out} = 200$ pc, $\dot{M}_{{\rm in}, r_{\rm in}}$ continues to increase even $2$ Myr into the run. 

The adiabatic index of the IC is changed from $\gamma_{\rm init} = 1.4$ to $1.6667$ from Run 17 to 18, 
and in the later runs $\gamma_{\rm init} = 1.6667$ is used.       
This change in $\gamma_{\rm init}$ has a relatively small impact on the mass inflow rate vs.~time 
(as can be seen from the bottom-left panel of Figure~\ref{fig-HC-Mdot_Rout200pc}), 
with a larger $\gamma_{\rm init}$ producing slightly lower $\dot{M}_{{\rm in}, r_{\rm in}}$. 

Three runs (17, 18, 21) are done with $T_{\infty} = 10^7$ K, which is the same $T_{\infty}$ as our spherical Bondi runs, 
for which the theoretical Bondi accretion rate is $\dot{M}_{B} = 4.9 \times 10^{24}$ g/s. 
In these simulations, the mass inflow rate is orders of magnitude higher than $\dot{M}_{B}$, 
and reaches a maximum value of $80 \dot{M}_{B}$ in Runs 17 and 18, and $2000 \dot{M}_{B}$ in Run 21. 
The reason is that in these runs 
the gas temperature is calculated self-consistently including adiabatic and radiative heating-cooling mechanisms, 
which reaches a value of $T (r_{\rm out}) \sim 10^5$\,K near the outer boundary 
(shown in Figure~\ref{fig-HC-Evol-T_Xi-T_r} and discussed in \S\ref{sec-results-HC-Thermal}). 
This outer temperature $T (r_{\rm out}) \sim 10^5$ K is lower than $T_{\infty} = 10^7$ K, 
and implies a larger mass accretion rate for the corresponding Bondi solution that the simulation is relaxing to. 

Therefore we do the remaining simulations with $T_{\infty} = 10^5$ K in Table~\ref{Table-HeatCool}. 
Runs 19 and 20 have $\rho_{\infty} = 10^{-21}$ g/cm$^3$, 
where the Bondi accretion rate is $\dot{M}_{B} = 4.9 \times 10^{27}$ g/s. 
As the top-left panel of Figure~\ref{fig-HC-Mdot_Rout200pc} indicates, the runs reach 
almost steady-state when considering time intervals of few $\times 10^4$ yrs, 
having an inflow rate between $1$ and $2~\dot{M}_{B}$. 
A larger outer radius $r_{\rm out} = 200$\,pc produces an increasing $\dot{M}_{{\rm in}, r_{\rm in}}$ for $0.7$ Myr (shown in top-right panel), 
reaching a maximum $6 \dot{M}_{B}$. 

Runs 22 and 23 have the same simulation parameters, 
but a different particle numbers of $N = 128^3$ and $256^3$. 
With a lower resolution, Run 22 is continued until the Bondi problem is represented correctly, 
or, as long as the mass inflow rate increases (\S\ref{sec-results-Bondi-Inflow}), 
reaching a maximum $\dot{M}_{{\rm in}, r_{\rm in}} = 5 \dot{M}_{B}$ after $\sim 3$\,Myr. 

Comparing the $T$ vs.~$r$ plots of Runs 22 and 23, we determine the dependence of artificial heating on resolution. 
We find that the temperature up to which the gas is artificially heated 
($T \sim 10^8$ K at $r_{\rm in}$) remains the same on changing resolution. 
However, the radius within which the extra heating occurs increases with lower resolution; 
in Run 22 gas within $\sim 1$ pc is heated, which is  $\sim 0.6$ pc in Run 23. 
This implies that a greater volume of gas is being affected by AV heating as resolution degrades. 

Reducing the initial temperatures to $T_{\rm init} = T_{\rm rad}$ brought a noteworthy change in our results: 
a larger mass fraction is accreted inside the inner radius rather than flowing outside the outer boundary. 
In the relevant runs 
$90 - 99 \%$ of the total initial mass accretes into $r_{\rm in}$, and only $1 - 5 \%$ of the total mass moves out of $r_{\rm out}$. 
With radiative heating and cooling incorporated, 
the gas temperature near the outer radius is between $T (r_{\rm out}) \sim 10^4 - 10^5$ K, 
which reduces the gas pressure w.r.t.~our previous runs which had $T (r_{\rm out}) = T_{\infty} = 10^7$ K. 
This reduced outward pressure gradient 
lowers the unwanted mass outflow at $r_{\rm out}$, and increases the desired inflow.

\subsubsection{The Thermal History of Particles} 
\label{sec-results-HC-Thermal} 

In Figure~\ref{fig-HC-Evol-T_Xi-T_r} we plot the time evolution of particles in the $T - \xi$ and $T - r$ planes of Run 18. 
Note that in the $T - \xi$ plane      
the right-most region (highest-$\xi$) is near the inner radius, 
the middle portion is the middle volume, and the left region is near the outer boundary. 
Initially the gas is in radiative equilibrium, 
therefore the particles lie on the $T_{\rm rad} - \xi$ curve at $t = 0$, 
where the inner particles has a temperature $T \sim 10^7$ K. 
The next epoch plotted is $t = 0.004$ Myr, and after that in intervals of $0.04$ Myr. 

In the $T - r$ plane, with the progress of time up to the epoch $t = 0.12$ Myr, 
more and more particles at $r < 10$\,pc are cooled to $T = 10^4 - 10^5$ K by mostly radiative cooling. 
There is a non-radiative heating near $r_{\rm in}$ causing particle temperature to rise to $T = 10^7 - 10^8$ K. 
The radius inside which this heating occurs decreases with time; particles inside $r = 0.6$\,pc are heated at $t = 0.04$ Myr, 
which reduces to heating inside $r = 0.15$\,pc at $t = 0.12$ Myr. 
We also see some heating at $r > 10$\,pc increasing in prominence with time. 
This is radiative heating because of the decreasing density near the outer boundary caused by the outflow. 
There is a cooling very close to $r = r_{\rm out} = 20$\,pc caused by adiabatic expansion, 
more prominent in the later epochs. 

The above features near $r_{\rm out}$ manifest themselves in the $T - \xi$ plane as the portions below the equilibrium curve, stretched out in $\xi$. 
Particles near the outer computational volume evolve along the $T_{\rm rad} - \xi$ curve from lower left to upper right as long as radiative processes dominate. 
Then they move out of the curve and stay at lower $T$, 
when radiative and adiabatic terms become comparable, and finally adiabatic cooling dominates at $r_{\rm out}$. 

By the last epoch shown in Figure~\ref{fig-HC-Evol-T_Xi-T_r}, $t = 0.16$ Myr, the simulation has 
started to deviate from the Bondi solution, 
because the mass inflow rate of Run 18 (Figure~\ref{fig-HC-Mdot_Rout200pc}, bottom-left panel) increases only up to $0.13$ Myr. 
The density of the whole volume at $t = 0.16$ Myr is lower than the Bondi profile by an order of magnitude. 
Therefore the photoionization parameter is higher, and the gas is heated up 
more than the temperature in the previous epochs uniformly at all radii. 


Figure~\ref{fig-HC-T_xi_Tinfinity} shows the temperature vs.~photoionization parameter of four runs in Table~\ref{Table-HeatCool}   
which has $T_{\rm init} = T_{\infty} = 10^7$ K, along with the $T - \xi$ equilibrium curve marked as the solid line. 
The gas in these runs is highly photoionized ($10^3 < \xi < 10^{13}$). 
The $T$ vs.~$\xi$ of nine other runs (having $T_{\rm init} = T_{\rm rad}$) in Table~\ref{Table-HeatCool} are shown in Figure~\ref{fig-HC-T_xi_Tradiative}, 
where the gas has a lower photoionization ($10^{-2} < \xi < 10^{5}$). 
Each run has a subset of particles falling on the $T - \xi$ equilibrium curve, 
implying that those are dominated by radiative processes. 
The increase in $T$ over $T_{\rm rad}$ at large-$\xi$ (upper-right in the $T - \xi$ plane) is due to non-radiative heating dominating near $r_{\rm in}$. 
The drop in $T$ below $T_{\rm rad}$ and the scatter at the left in Figure~\ref{fig-HC-T_xi_Tinfinity}, 
and that at intermediate-$\xi$ (lower-right in the $T - \xi$ plane) in Figure~\ref{fig-HC-T_xi_Tradiative} 
is caused by adiabatic cooling due to the outflow at $r_{\rm out}$. 

The non-radiative heating toward $r_{\rm in}$ seen in Figs.~\ref{fig-HC-Evol-T_Xi-T_r}, \ref{fig-HC-T_xi_Tinfinity} and \ref{fig-HC-T_xi_Tradiative} 
is caused by a combination of adiabatic compression and numerical AV near the inner radius.

\subsubsection{Dependence on X-ray Luminosity}    
\label{sec-results-HC-Lx} 

We explore the dependence of gas accretion properties on the X-ray luminosity of the corona by performing runs with different $L_{X}$. 
Runs 15 and 16 are done with $L_{X} / L_{\rm Edd} = 0.5$ and $5 \times 10^{-4}$, 
whose mass inflow rates are shown in the bottom panel of Figure~\ref{fig-HC-Mdot_Rout_LX}. 
These runs have $\rho_{\infty} = 10^{-27}$ g/cm$^3$, corresponding to a Bondi rate of $\dot{M}_{B} = 4.9 \times 10^{18}$ g/s. 
The run with the lower $L_{X}$ produces a higher inflow rate. 
Between $t = (1.5 - 2.5) \times 10^4$ yrs, when the runs are close to steady-state, 
$\dot{M}_{{\rm in}, r_{\rm in}}$ is $1$ and $0.8$ times $\dot{M}_{B}$ for $L_{X} / L_{\rm Edd} = 5 \times 10^{-4}$ and $0.5$, respectively. 
A higher X-ray luminosity corona heats up the gas more, causes the gas to expand, 
hence increases the outflow rate at $r_{\rm out}$ and decreases the inflow rate at $r_{\rm in}$. 
But the reduction in $\dot{M}_{{\rm in}, r_{\rm in}}$ is  
smaller only by a factor of $0.8$ when $L_{X}$ is increased by $10^3$ times, in this particular case. 
This is because of the very low gas density of these runs, 
which causes the gas to be highly photoionized with these $L_{X}$ (Figure~\ref{fig-HC-T_xi_Tinfinity}). 
The average photoionization parameter decreases from $\xi (L_{X} = 0.5) \sim 10^{11}$ 
to $\xi (L_{X} = 5 \times 10^{-4}) \sim 10^{8}$, 
which is still high, and the gas remains around the `hot-branch' of the $T - \xi$ radiative equilibrium curve. 
The low-density gas is almost equally heated up with these values of $L_{X}$. 
Consequently, we also see that decreasing $L_{X} / L_{\rm Edd}$ from $0.5$ to $5 \times 10^{-4}$ 
has no effect on the outflow ($\sim 0.98$) and inflow ($\sim 0.005$) mass fractions in these simulations. 


We investigate more varying-$L_{X}$ cases by restarting Run 23 (with $L_{X} / L_{\rm Edd} = 5 \times 10^{-4}$) at $t = 1.4$ Myr, 
and performing subsequent Runs $24 - 28$ with: 
$L_{X} / L_{\rm Edd} = 5 \times 10^{-5}, 5 \times 10^{-3}, 0.01, 0.02, 0.05$. 
With larger $L_{X}$, particles have higher photoionization parameter, 
and shifts to the right of the $T - \xi$ equilibrium curve. 
This can be seen by comparing the $T - \xi$ planes of the six runs in Figure~\ref{fig-HC-T_xi_Tradiative} 
in lower two panels; 
particles in Run 24 (lowest $L_{X}$) occupies the left-most area, 
in Runs $23 - 28$ they progressively shift right as $L_{X}$ increases. 
Also with rising $L_{X}$, particles are in radiative equilibrium up to a higher temperature: 
$T \sim 10^6$ K in Run 25, up to $T \sim T_C = 3 \times 10^7$ K (the Compton equilibrium temperature) in Run 28, 
because radiative heating dominates with such X-ray luminosity. 

The mass inflow rate at the inner radius of Runs $23 - 28$ are shown in Figure~\ref{fig-HC-Mdot_LXray}. 
We see that $\dot{M}_{{\rm in}, r_{\rm in}}$ drops at a varying rate as $L_{X}$ is increased, because the gas is heated up and expands. 
There is only a slight difference in the mass inflow rate by changing $L_{X}$ from $5 \times 10^{-5}$ to $5 \times 10^{-3}$. 
The two lowest $L_{X}$ runs ($24$ and $23$) have the same $\dot{M}_{{\rm in}, r_{\rm in}}$, 
and $L_{X} / L_{\rm Edd} = 5 \times 10^{-3}$ (Run 25) produces $0.95$ times lower $\dot{M}_{{\rm in}, r_{\rm in}}$. 
The central inflow rate becomes $0.8$ times lower with $L_{X} / L_{\rm Edd} = 0.01$ in Run 26. 

Further increases of the value of $L_{X}$ by a factor of $2$ or more show significant changes in the gas dynamical behavior. 
The mass inflow rate decreases by $3 - 4$ orders of magnitude, 
and an outflow develops because the gas is heated up by the higher X-ray luminosity. 
We see that the transition from net inflow to net outflow occurs 
with a value of $L_{X} / L_{\rm Edd}$ between $0.01$ and $0.02$, 
for $\rho_{\infty} = 10^{-23}$ g/cm$^3$ and $T_{\infty} = 10^5$ K. 
The mass inflow rate in Run 27 rises again after $2.2$ Myr,  
because the outflow is not strong enough with $L_{X} / L_{\rm Edd} = 0.02$, and the inflow resumes later. 
There is however a strong outflow with $L_{X} / L_{\rm Edd} = 0.05$ which expels a substantial fraction of gas 
out of the computational volume, and $\dot{M}_{{\rm in}, r_{\rm in}}$ never rises within $5$ Myr. 

We find that, in Runs $26$ and $27$, the gas cools and gets denser non-spherically leading to fragmentation after a time of $\sim 1.6$ Myr. 
Then the mass inflow rate in these runs becomes very noisy (Figure~\ref{fig-HC-Mdot_LXray}). 
This fragmentation appears to be related to a growth of thermal instability. 
However, numerical effects might also be of some relevance which we will investigate in the future. 

The temperature profile as a function of radius for the three lower-$L_{X}$ runs are shown in Figure~\ref{fig-HC-T_r_3Lx}. 
With $L_{X} / L_{\rm Edd} = 5 \times 10^{-5}$ and $5 \times 10^{-4}$, particles in most of the volume have a low temperature $T < 10^5$ K. 
There is a significant heating at $r < 0.2$\,pc in Run 24, and at $r < 0.6$\,pc in Run 23, increasing temperatures to $T \sim 10^7 - 10^8$ K. 
With $L_{X} / L_{\rm Edd} = 5 \times 10^{-3}$, particles at all radii are heated 
with an almost constant slope. 
As $L_{X}$ increases, $\xi$ rises and the particles reach a higher radiative equilibrium temperature, 
so the whole $T - r$ curve shifts upward in $T$, visible in the panels from left to right in Figure~\ref{fig-HC-T_r_3Lx}. 
This higher $T_{\rm rad}$ and the inner heating 
combined together has an indistinguishable slope in the $T - r$ plane of Run 25. 

Indications of AV heating toward $r_{\rm in}$ can be seen in Figure~\ref{fig-HC-T_r_3Lx}. 
The predicted temperature profiles $T_{ff,ar}$ and $T_{ff, a}$ (\S\ref{sec-results-HC-CompareBondi}) 
are plotted as the solid and dashed curves in each panel. 
The simulation particles near the inner radius are heated above both the predictions. 
It is more prominent in Runs 24 and 23 (left and middle panels), but all the three runs show it. 
The radius inside which the AV heating occurs depends on $L_{X}$. 
Gas inside $r \leq 0.2$\,pc is artificially heated with $L_{X} / L_{\rm Edd} = 5 \times 10^{-5}$, 
and the gas inside $r \leq 0.5$\,pc for $L_{X} / L_{\rm Edd} \geq 5 \times 10^{-4}$.

The gas properties as a function of radius in Run 26 ($L_{X} / L_{\rm Edd} = 0.01$) is plotted in Figure~\ref{fig-HC-ScatterVsR-Lx1e-2}, 
where the gas is radiated by $20$ times higher X-ray luminosity as compared to 
Run 23 ($L_{X} / L_{\rm Edd} = 5 \times 10^{-4}$, Figure~\ref{fig-HC-ScatterVsR}). 
The radial velocity in Run 26 is slower than free-fall, but the density profile is qualitatively similar to that in Run 23. 
The temperature profile shows radiative heating by a higher X-ray luminosity, up to the inner boundary. 
The Mach number reduces, because the gas becomes hotter increasing the sound speed, 
and the gas is mildly super-sonic (Mach $\sim 1.5 - 2$) as it accretes onto $r_{\rm in}$. 

Figure~\ref{fig-HC-ScatterVsR-Lx5e-2} shows the radial gas properties of Run 28 ($L_{X} / L_{\rm Edd} = 0.05$), 
where the gas is violently heated and expands.   
There is a well-formed outflow between $r \sim 3 - 25$ pc, where all the particles have $v_r > 0$. 
This is to be distinguished from the outflow near $r_{\rm out}$ owing to the outer boundary condition. 
The relevant outflow in Run 28 occurs at intermediate $r$ values, 
and is surrounded by a region of inflowing gas between $r \sim 25 - 150$ pc. 
So it is clearly the previously free-fall inflowing gas in Run 23 that has reacted to the increased $L_{X}$, 
has slowed down, and is eventually outflowing at the time plotted ($1.5$ Myr). 
This outflow is caused by the X-ray heating of the gas due to the high $L_{X}$ value in Run 28. 

The intensity of the heating can also be seen from the scarcity of particles at $r < 0.7$\,pc in Figure~\ref{fig-HC-ScatterVsR-Lx5e-2}. 
The gas is heated and expands, and the rate of inflow from the outer volume has reduced. 
This causes a decrease in density near $r_{\rm in}$, about $50$ times lower than Run 26. 
The gas becomes sub-sonic almost throughout the volume, except Mach $\sim 1$ for few accreting particles next to $r_{\rm in}$. 
The temperature profile shows radiative heating at all radii. 
The small temperature spike at $r \sim 25$\,pc corresponds to the heating caused by 
the transition from outflowing gas inside to inflowing gas outside $25$ pc. 

Summarising the effect of X-ray luminosity of the central corona,   
we find that a higher $L_X$ produces a lower mass inflow rate at the inner boundary. 
This is because of feedback due to radiative heating occurring in our simulations. 
With high enough value of $L_X$, the gas thermally expands.


We perform some 1-D and 2-D simulations with the ZEUS code for comparison. 
These calculations confirm some of the results we obtained in our GADGET runs. 
In the ZEUS counterpart of Run 23, 
the mass inflow rate at the inner boundary settles at the value of $5 \times 10^{25}$ g/s after a time of $8$ Myr. 
This time duration is $4$ times longer than the duration for which the GADGET Run 23 is evolved. 
To continue the GADGET run for a comparable time, significant computational resources would be required. 
ZEUS counterparts of simulation Runs 26 and 27 also show central mass inflow rates consistent with our GADGET results.

\subsubsection{Controlling Artificial Viscosity Heating}    
\label{sec-results-HC-AV} 

The standard AV prescription used in SPH \citep{Monaghan83} 
is implemented in the GADGET code \citep[see][Eqs.~$9-17$]{Springel05}. 
The velocity of particles is reduced by a viscous force, which generates extra entropy. 
This transforms kinetic energy of gas motion irreversibly into heat. 
The parametrization of AV in GADGET is taken from \citet{Monaghan97}. 
All our simulations are done with a viscosity parameter $\alpha = 0.8$ (values between $\alpha \simeq 0.5 - 1.0$ are typical). 

This AV heating is a numerical artifact in some of our simulations, 
which increases the gas temperature higher than adiabatic and radiative predictions. 
It could be present in any other studies using SPH, and without proper testing it 
could lead to wrong conclusions about heating mechanisms. 
In our simulations, 
AV does not affect the mass inflow rate at the inner boundary since only the highly supersonic parts of the inflow are heated. 
But one must be careful if using the properties of this flow to 
investigate gas behavior on scales smaller than the inner radius ($0.1$ pc), 
and infer some implications for processes like star formation and the formation of an accretion disk. 

In certain simulations it might be possible to 
control the magnitude of AV, or try an alternate AV scheme to minimize such undesired dissipation. 
In general, however, the adverse effect of AV is unknown in an SPH simulation, 
making it hard to disentangle its effects from other known mechanisms. 
\citet{Thacker00} found that AV is the single most important factor distinguishing the results when they analysed twelve different implementations of SPH in cosmological simulations. 

Artificial viscosity is needed in SPH 
to prevent inter-penetration of SPH particles and thus a multi-valued flow, to allow shocks to form, and to damp post shock oscillations. 
However, AV might cause unphysical dissipation away from shocks and unwanted heating. 
In order to reduce these problems, various modifications to the standard AV recipe in SPH have been proposed: 
the switch by \citet{Balsara95}, 
using time-varying viscosity coefficient for each particle \citep{Morris97}, 
restricting the von Neumann-Richtmeyer AV to only non-compressed gas 
and calculating the bulk AV from an integrated mean velocity in a particle's neighbourhood \citep{Selhammar97}, 
using the total time derivative of the velocity divergence as a shock indicator and adapting individual viscosities for particles \citep{Cullen10}. 
Other dissipation-inhibiting modified AV prescriptions have been reviewed in \citet{Monaghan05} and \citet[][and references therein]{Rosswog09}. 

A few tests of modified AV schemes in certain astrophysical scenarios show that 
the proposed improved AV recipes are not always effective. 
\citet{Cartwright10} tested the Balsara switch and time dependent viscosity 
for their effectiveness in disabling AV in non-convergent Keplerian shear flow, in SPH simulations of accretion disks.  
They found that neither of the mechanisms work effectively, as those methods leave AV active in areas of smooth shearing flow. 
\citet{Junk10} showed that the Kelvin-Helmholtz instability is increasingly suppressed for shear flows with rising density contrasts, 
where using the Balsara switch does not solve the problem, 
indicating that other changes to the SPH formalism are required in order to correctly model shearing layers of different densities. 

In a future study we will test if some alternate AV prescription can reduce the excessive inner heating 
occurring in our spherical accretion simulations.

\section{Summary and Conclusion} 
\label{sec-conclusion} 

Aiming to answer the question how well can the SPH technique follow spherical Bondi accretion, 
we perform numerical simulations of the Bondi problem using the 3D SPH code GADGET. 
Our simulation setup comprise of a spherical distribution of gas, between the length scale $0.1 - 200$ pc, accreting onto a central SMBH. 
Represented by discrete particles in SPH, the gas undergoes hydrodynamic interactions, 
and experiences gravitational potential of the BH, implemented by the Paczynsky-Wiita `static' potential. 
The values of model parameters we explore are selected in order to 
encompass the Bondi and sonic radii well within our computational volume. 

Our spherical accretion simulations (Table~\ref{Table-Bondi}, $\gamma_{\rm run} =  1.01$) 
can reproduce Bondi accretion within limited temporal and spatial domains. 
The main results are summarised below (points $1 - 4$): 

(1) The radial profiles of the velocity and density of the simulation particles match the 
corresponding Bondi solutions quite well, except near the inner and outer radii. 
The temperature profile is almost isothermal at $10^7$ K (= $T_{\infty}$). 
The gas is subsonic near $r_{\rm out}$, crosses a sonic point at $\sim 1.5$\,pc 
(consistent with the predicted $R_s$ value of the Bondi solution), and approaches $r_{\rm in}$ supersonically. 

The deviations of the profiles from the ideal Bondi solution at small radii is caused by three effects:   
numerical resolution (smoothing length of the SPH particles), 
no special corrections to the inner boundary conditions \citep{Bate95}, and artificial viscosity in the SPH code we are using. 
The deviations at the outer boundary is caused by the outflow of particles, occurring in several of our simulations, 
because of the outward pressure gradient at $r_{\rm out}$. 

(2) Runs starting with the Bondi IC reach steady-state quickly (earlier than starting with a uniform IC). 
The mass inflow rate at the inner boundary is equal to the Bondi mass accretion rate 
($\dot{M}_{B} = 4.6 \times 10^{27}$ g/s) 
for time durations of few $\times 10^{4}$ yrs, amounting to few times the Bondi time. 
The duration for which Bondi accretion is reasonably reproduced 
lengthens with increasing outer radius of the computational volume. 
On adding an external galaxy bulge potential to the simulation, the mass inflow rate increases w.r.t.~the Bondi rate,  
reaching a maximum value of $1.6 \dot{M}_{B}$. 

(3) The gas in our simulations accretes at the Bondi rate in the inner parts,     
between $r_{\rm in}$ and $r_{tr}$ (the transition radius from net inflow inside $r_{tr}$ to net outflow outside it, \S\ref{sec-results-Bondi-MassFlux}), 
but an outflow develops and dominates in the outer regions. 
We see that $r_{tr} = r_{\rm out}$ (i.e., the whole computational volume is accreting at the Bondi rate) only at very early simulation epochs. 
Later on, $r_{tr}$ reduces and starts to move inward. 


(4) More stringent boundary conditions can reduce the outflow at $r_{\rm out}$ to a negligible fraction. 
Two of our tests were successful, 
where we change the velocity and pressure gradient (one in each case) of every particle having a net outward radial velocity ($v_r > 0$)
in a wide outer shell, between $[0.4 r_{\rm out} - r_{\rm out}]$.  
With such a boundary condition, 
the mass inflow rate is nearly-steady ($\dot{M}_{{\rm in}, r_{\rm in}} \simeq \dot{M}_{B}$) for durations up to $\sim 25 t_B$.

\vspace{0.5cm}

Next, we include radiative heating by a central X-ray corona and radiative cooling in our next series of simulations 
(Table~\ref{Table-HeatCool}, $\gamma_{\rm run} =  5/3$), whose results are in the following (points $5 - 9$):

(5) The inner computational volume ($r < 10$ pc) is inflow-dominated 
for relatively-low values of the X-ray luminosity ($L_{X} / L_{\rm Edd} = 5 \times 10^{-5} - 5 \times 10^{-3}$). 
In these simulations 
the particle velocity and density match the theoretical free-fall scalings well in the middle volume, 
with a deviation near the inner and outer radii, for the same reasons mentioned in point (1) before. 
The Mach number increases inward up to a certain $r$, but after that decreases at smaller $r$ (because of a strong inner heating). 

(6) The runs with radiative heating and cooling take longer times to reach a steady-state, compared to the Bondi accretion runs. 
Some of the runs reach almost steady-state, when considering time intervals of few $\times 10^4$ yrs.   

(7) In each run, some particles are in perfect radiative equilibrium.    
Some particles are above or below the $T - \xi$ equilibrium curve, and are dominated by non-radiative heating and cooling, respectively. 
We identified three non-radiative mechanisms: 
heating near $r_{\rm in}$ (at the highest-$\xi$, or lowest-$r$ in a run) caused by adiabatic compression and numerical artificial viscosity, 
and adiabatic cooling near $r_{\rm out}$ because of the outflow expansion at the outer boundary. 

(8) We find the occurrence of a heating of the inflowing gas in the inner volume, which is dominated by artificial viscosity. 
This is a powerful heating causing the temperature to rise to $T \sim 10^7 - 10^8$ K, 
which is orders of magnitude above radiative equilibrium in some runs (Figure~\ref{fig-HC-T_xi_Tradiative}). 
Standard AV prescriptions in SPH generates extra entropy at the expense of kinetic energy, 
therefore the velocity of the particles is smaller than the theoretical free-fall prediction at the same $r$ in our simulations. 
The radius inside which the excessive AV heating occurs 
depends on the simulation parameters; we see it at $r < 0.2 - 0.8$\,pc in different runs. 

(9) We detect signatures of 
radiative feedback between the characteristic X-ray luminosity of the central spherical corona 
(the radiation source, \S\ref{sec-numerical-RadHC}) and the mass inflow rate at the inner boundary in our simulations. 
We see that the central inflow decreases at a varying rate as $L_{X}$ is increased. 
High X-ray luminosities heat up the gas more, causing it to expand, increasing the outflow at $r_{\rm out}$, 
and decreasing the inward flow at $r_{\rm in}$. 
With high enough values of $L_X$, the inflow at $r_{\rm in}$ slows down, and an outflow develops in the inner volume. 
We see the outflow occurring for values of $L_{X} / L_{\rm Edd} \geq 0.02$, when $\rho_{\infty} = 10^{-23}$ g/cm$^3$. 

We find non-spherical fragmentation and clumping of the gas in the runs with $L_{X} / L_{\rm Edd} = 0.01$ and $0.02$. 
This is likely due to a development of thermal instability 
which is triggered by small, non-spherical  fluctuations in the flow properties inherent to SPH. 
We will present results of our detailed analysis of the clumping in a future publication.

\section*{Acknowledgments} 
PB thanks J.-H. Choi for helpful conversations, and we are grateful to V. Springel for allowing us to use the GADGET-3 code. 
This work is supported in part by the NSF grant AST-0807491, National Aeronautics and Space Administration under Grant/Cooperative Agreement No. NNX08AE57A issued by the Nevada NASA EPSCoR program, and the President's Infrastructure Award from UNLV.
This research is also supported by the NSF through the TeraGrid resources provided by the Texas Advanced Computing Center.
Some numerical simulations and analyses have also been performed on the UNLV Cosmology Cluster.

%

\appendix

\section{Solving the Specific Internal Energy Equation}
\label{sec-appenix} 

We work out simple solutions of the energy equation including adiabatic and radiative terms, given radially-dependent velocity and density profiles. 
The hydrodynamic equation for the specific internal energy $u$ is, 
\begin{equation} 
\label{eq-dudt} 
\frac{Du} {Dt} = \left( \frac{\partial}{\partial t} + \vec{v} \cdot \vec{\nabla} \right) u
= - \frac{P}{\rho} \left( \vec{\nabla} \cdot \vec{v} \right) + {\cal{L}} ,
\end{equation} 
where, the first term in the RHS is adiabatic and the second is radiative. 
In terms of the temperature $T = \mu (\gamma - 1) u / k$, 
assuming steady-state (i.e., the time derivative is 0) $\partial T/ \partial t = 0$, for a radial velocity $\vec{v} = v_r \hat{r}$, 
the energy equation can be rewritten as, 
\begin{equation} 
\frac{\partial T} {\partial r} = 
\frac{ \left(\gamma - 1\right) } {v_r} \left[ - \frac{T}{r^2} \frac{\partial}{\partial r} \left( r^2 v_r \right) + \frac{\mu}{k}  {\cal{L}} \right] . 
\end{equation} 
For a free-fall velocity (Eq.~\ref{eq-Vff}), the energy equation reduces to, 
\begin{equation} 
\label{eq-dTdr_vff} 
\frac{\partial T} {\partial r} = 
\left(\gamma - 1\right) \left[ - \frac{3 T}{2 r} + \frac{ \mu {\cal{L}} } {k  v_{ff}} \right] . 
\end{equation} 

We solve Eq.~(\ref{eq-dTdr_vff}) numerically, to find $T$ as a function of $r$, 
using a fully explicit integration method with small enough $\Delta r$ steps to prevent negative temperatures ($\Delta r = 0.002 - 0.0002$ pc), 
\begin{equation} 
T \left( r + \Delta r \right) = T \left( r \right) + \frac{\partial T} {\partial r} \left( r \right)  \Delta r . 
\end{equation} 
An initial value of $T = 2 \times 10^4$ K at $r = r_{\rm out} = 200$\,pc is used during integration. 
We take the gas velocity as free-fall (Eq.~\ref{eq-Vff}), 
and the free-fall density (Eq.~\ref{eq-Rho_ff}) assuming appropriate scaling from the particle snapshot plot. 

\end{twocolumn}


\begin{onecolumn}

%

\begin{deluxetable}{cccccccccc} 
\tabletypesize{\scriptsize}
\tablewidth{0pt} 
\tablecaption{Simulations of Bondi Accretion \tablenotemark{a}} 
\tablehead{ 
\colhead{Run} & 
\colhead{$r_{\rm out}$} & 
\colhead{$N$ \tablenotemark{b}} & 
\colhead{IC} & 
\colhead{$M_{\rm tot,IC}$ \tablenotemark{c}} & 
\colhead{$M_{\rm part}$ \tablenotemark{d}} & 
\colhead{$t_{\rm end}$ \tablenotemark{e}} & 
\colhead{$f_{\rm IN}$ \tablenotemark{f}} & 
\colhead{$f_{\rm OUT}$ \tablenotemark{g}} & 
\colhead{$\dot{M}_{{\rm in}, r_{\rm in}}$ \tablenotemark{h}} 
\\ 
\colhead{No.} & 
\colhead{[pc]} & & & 
\colhead{$[ M_{\odot} ]$} & 
\colhead{$[ M_{\odot} ]$} & 
\colhead{[$10^{4}$ yr]} & & & 
\colhead{$[ \dot{M}_{B} (\gamma_{\rm run}, \rho_{\infty}, T_{\infty}) ]$} 
} 

\startdata 


$1$ & $5$ & $64^3$ & Uniform \tablenotemark{i} & $3.96 \times 10^{5}$ & $1.51$ & $3$ & $0.35$ & $0.65$ & $0.4$ \\   

$2$ & $10$ & $64^3$ & Uniform & $6.19 \times 10^{6}$ & $23.61$ & $7.2$ & $0.15$ & $0.85$ & $0.6$ \\ 

$3$ & $50$ & $128^3$ & Uniform & $7.73 \times 10^{8}$ & $368.60$ & $20$ & $0.02$ & $0.9$ & $1$ \\ 

\hline \\ 

$4$ & $5$ & $64^3$ & Bondi \tablenotemark{j} & $1.81 \times 10^{6}$ & $6.89$ & $2$ & $0.4$ & $0.55$ & $1$ \\ 

\hline \\ 

$5$ & $10$ & $64^3$ & Bondi & $9.76 \times 10^{6}$ & $37.23$ & $8$ & $0.2$ & $0.8$ & $1$ \\ 

$6$ & $10$ & $128^3$ & Bondi & $9.76 \times 10^{6}$ & $4.65$ & $8$ & $0.2$ & $0.8$ & $1$ \\ 


\hline \\ 

$7$ & $20$ & $128^3$ & Bondi & $6.24 \times 10^{7}$ & $29.75$ & $8$ & $0.07$ & $0.9$ & $1$ \\ 

\hline \\ 

$7a$ \tablenotemark{k} & $20$ & $128^3$ & Bondi & $6.24 \times 10^{7}$ & $29.75$ & $80$ & $0.65$ & $0.05$ & $1$ \\ 


$7b$ \tablenotemark{l} & $20$ & $128^3$ & Bondi & $6.24 \times 10^{7}$ & $29.75$ & $100$ & $0.93$ & $0.01$ & $1.2$ \\

\hline \\ 

$8$ & $50$ & $128^3$ & Bondi & $8.48 \times 10^{8}$ & $404.35$ & $16$ & $0.02$ & $0.85$ & $1$ \\ 

\hline \\ 

$9$ & $20$ & $128^3$ & $\rho_B, v_{\rm init} = 0$ & $6.24 \times 10^{7}$ & $29.75$ & $8$ & $0.06$ & $0.9$ & $1$ \\ 

$10$ & $20$ & $128^3$ & Uniform & $4.95 \times 10^{7}$ & $23.60$ & $8$ & $0.06$ & $0.9$ & $1$ \\ 

$11$ & $20$ & $128^3$ & Hernquist \tablenotemark{m} & $6.24 \times 10^{7}$ & $29.75$ & $7.2$ & $0.06$ & $0.85$ & $1.4$ \\ 

\hline \\ 

$12$ \tablenotemark{n} & $20$ & $128^3$ & Bondi & $6.24 \times 10^{7}$ & $29.75$ & $8$ & $0.08$ & $0.8$ & $1.6$ \\ 

\enddata 
\label{Table-Bondi}

\tablenotetext{a}{ All the runs have $r_{\rm in} = 0.1$\,pc and $\gamma_{\rm run} =  1.01$. 
Initial conditions are generated using 
$\gamma_{\rm init} = 1.01$, $\rho_{\infty} =  10^{-19}$ g/cm$^3$, $T_{\infty} = 10^7$ K, and $T_{\rm init} = T_{\infty}$. 
The corresponding Bondi solution has 
$R_B = 3.0$ pc, $R_s = 1.5$ pc, and $t_B = 7.9 \times 10^3$ yr. 
}
\tablenotetext{b}{ $N$ = Number of particles in the initial condition.}
\tablenotetext{c}{ $M_{\rm tot,IC} =$ Initial total gas mass within simulation volume.} 
\tablenotetext{d}{ $M_{\rm part} =$ Particle mass.} 
\tablenotetext{e}{ $t_{\rm end} =$ Simulation end time.}  
\tablenotetext{f}{ $f_{\rm IN} =$ Mass fraction accreted into $r_{\rm in}$ by $t = t_{\rm end}$.}  
\tablenotetext{g}{ $f_{\rm OUT} =$ Mass fraction moved outside $r_{\rm out}$ by $t = t_{\rm end}$.} 

\tablenotetext{h}{ The steady-state value of mass inflow rate at the inner radius (if a steady-state is reached in a run), 
or the maximum value (if a steady-state is not reached by $t_{\rm end}$).} 

\tablenotetext{i}{ Uniform initial condition: $\rho_{\rm init} = \rho_{\infty}, v_{\rm init} = 0$. } 
\tablenotetext{j}{ Bondi initial condition: $\rho_{\rm init} (r) = \rho_B (r), v_{\rm init} (r) = v_B (r)$. } 

\tablenotetext{k}{ Run with different outer boundary condition: Set $v_r = v_B$ for particles which have $v_r > 0$ in between $8 - 20$ pc.} 
\tablenotetext{l}{ Run with different outer boundary condition: 
Set zero pressure gradient, $\langle \nabla P / \rho \rangle_i = 0$, for all particles between $18 - 20$ pc 
and for those which have $v_r > 0$ in between $8 - 20$ pc.} 

\tablenotetext{m}{ Hernquist initial condition: $\rho_{\rm init} (r) = \rho_H (r), v_{\rm init} = 0$. } 

\tablenotetext{n}{ Run with a bulge potential of a Milky-Way type galaxy following the Hernquist profile (Eq.~\ref{eq-PsiH}) 
with $M_{\rm bulge} = 3.4 \times 10^{11} M_{\odot}$ and $a_{\rm bulge} = 700$ pc.} 

\end{deluxetable} 

%

\begin{deluxetable}{ccccccccccccccc} 
\rotate 
\tabletypesize{\scriptsize}
\tablewidth{0pt} 
\tablecaption{Simulations of Spherical Accretion with Radiative Heating and Cooling \tablenotemark{a}} 
\tablehead{ 
\colhead{Run} & 
\colhead{$r_{\rm out}$} & 
\colhead{$N$} & 
\colhead{$M_{\rm tot,IC}$} & 
\colhead{$M_{\rm part}$} & 
\colhead{$\gamma_{\rm init}$} & 
\colhead{$T_{\infty}$} & 
\colhead{$R_B$} & 
\colhead{$\rho_{\infty}$} & 
\colhead{$T_{\rm init}$} & 
\colhead{$L_{X}$} & 
\colhead{$t_{\rm end}$} & 
\colhead{$f_{\rm IN}$} & 
\colhead{$f_{\rm OUT}$} & 
\colhead{$\dot{M}_{{\rm in}, r_{\rm in}}$} 
\\ 
\colhead{No.} & 
\colhead{[pc]} & & 
\colhead{$[ M_{\odot} ]$} & 
\colhead{$[ M_{\odot} ]$} & & 
\colhead{[K]} & 
\colhead{[pc]} & 
\colhead{[g/cm$^3$]} & & 
\colhead{[$L_{\rm Edd}$]} & 
\colhead{[$10^5$ yr]} & & & 
\colhead{$[ \dot{M}_{B} (\gamma_{\rm run}, \rho_{\infty}, T_{\infty}) ]$} 
} 

\startdata 


$13$ & $20$ & $128^3$ & $5.81 \times 10^{5}$ & $0.277$ & $1.4$ & $10^7$ & $2.19$ & $10^{-21}$ & $T_{\infty}$ & $0.5$ & $1.0$ & $0.005$ & $0.99$ & $2$ \\ 

$14$ & $50$ & $128^3$ & $8.23 \times 10^{6}$ & $3.92$ & $1.4$ & $10^7$ & $2.19$ & $10^{-21}$ & $T_{\infty}$ & $0.5$ & $2.9$ & $0.001$ & $0.99$ & $1.2$ \\ 

\hline \\ 

$15$ & $20$ & $128^3$ & $5.81 \times 10^{-1}$ & $2.77 \times 10^{-7}$ & $1.4$ & $10^7$ & $2.19$ & $10^{-27}$ & $T_{\infty}$ & $0.5$ & $1.0$ & $0.005$ & $0.99$ & $2$ \\


$16$ & $20$ & $256^3$ & $5.81 \times 10^{-1}$ & $3.46 \times 10^{-8}$ & $1.4$ & $10^7$ & $2.19$ & $10^{-27}$ & $T_{\infty}$ & $5 \times 10^{-4}$ & $1.9$ & $0.0052$ & $0.98$ & $2$ \\ 

\hline \\ 

$17$ & $20$ & $128^3$ & $5.81 \times 10^{5}$ & $0.277$ & $1.4$ & $10^7$ & $2.19$ & $10^{-21}$ & $T_{\rm rad}$ \tablenotemark{b} & $5 \times 10^{-4}$ & $2.9$ & $0.97$ & $0.03$ & $80$ \\ 

$18$ & $20$ & $128^3$ & $5.65 \times 10^{5}$ & $0.269$ & $5/3$ & $10^7$ & $1.84$ & $10^{-21}$ & $T_{\rm rad}$ & $5 \times 10^{-4}$ & $3.0$ & $0.97$ & $0.03$ & $80$ \\ 

$19$ & $20$ & $128^3$ & $1.47 \times 10^{7}$ & $7.0$ & $5/3$ & $10^5$ & $183.9$ & $10^{-21}$ & $T_{\rm rad}$ & $5 \times 10^{-4}$ & $1.5$ & $0.99$ & $0$ & $2$ \\ 

\hline \\ 

$20$ & $200$ & $256^3$ & $1.33 \times 10^{9}$ & $79.09$ & $5/3$ & $10^5$ & $183.9$ & $10^{-21}$ & $T_{\rm rad}$ & $5 \times 10^{-4}$ & $6.5$ & $0.13$ & $0$ & $6$ \\ 

$21$ & $200$ & $256^3$ & $4.95 \times 10^{8}$ & $29.50$ & $5/3$ & $10^7$ & $1.84$ & $10^{-21}$ & $T_{\rm rad}$ & $5 \times 10^{-4}$ & $8.7$ & $0.1$ & $0.012$ & $2000$ \\ 

\hline \\ 

$22$ & $200$ & $128^3$ & $1.33 \times 10^{7}$ & $6.33$ & $5/3$ & $10^5$ & $183.9$ & $10^{-23}$ & $T_{\rm rad}$ & $5 \times 10^{-4}$ & $70$ & $0.9$ & $0.1$ & $5$ \\ 

$23$ & $200$ & $256^3$ & $1.33 \times 10^{7}$ & $0.791$ & $5/3$ & $10^5$ & $183.9$ & $10^{-23}$ & $T_{\rm rad}$ & $5 \times 10^{-4}$ & $20$ & $0.3$ & $0.07$ & $4$ \\ 

\hline \\ 

$24$ \tablenotemark{c} & $200$ & $1.24 \times 10^{7}$ & $9.77 \times 10^{6}$ & $0.791$ & $5/3$ & $10^5$ & $183.9$ & $10^{-23}$ & $T_{\rm Run23}$ & $5 \times 10^{-5}$ & $19$ & $0.12$ & $0.01$ & $4$ \\    

$25$ & $200$ & $1.24 \times 10^{7}$ & $9.77 \times 10^{6}$ & $0.791$ & $5/3$ & $10^5$ & $183.9$ & $10^{-23}$ & $T_{\rm Run23}$ & $5 \times 10^{-3}$ & $21$ & $0.15$ & $0.02$ & $3.8$ \\    


$26$ & $200$ & $1.24 \times 10^{7}$ & $9.77 \times 10^{6}$ & $0.791$ & $5/3$ & $10^5$ & $183.9$ & $10^{-23}$ & $T_{\rm Run23}$ & $1 \times 10^{-2}$ & $22$ & $0.2$ & $0.05$ & $3.4$ \\ 

$27$ & $200$ & $1.24 \times 10^{7}$ & $9.77 \times 10^{6}$ & $0.791$ & $5/3$ & $10^5$ & $183.9$ & $10^{-23}$ & $T_{\rm Run23}$ & $2 \times 10^{-2}$ & $25$ & $0.03$ & $0.1$ & $0.002$ \\ 

$28$ & $200$ & $1.24 \times 10^{7}$ & $9.77 \times 10^{6}$ & $0.791$ & $5/3$ & $10^5$ & $183.9$ & $10^{-23}$ & $T_{\rm Run23}$ & $5 \times 10^{-2}$ & $50$ & $0.008$ & $0.99$ & $0.0001$ \\  

\enddata 
\label{Table-HeatCool} 

\tablenotetext{a}{ All the runs have $r_{\rm in} = 0.1$\,pc and $\gamma_{\rm run} = 5/3$. 
Initial conditions are generated using $\rho_{\rm init} (r) = \rho_B (r)$, and $v_{\rm init} (r) = v_B (r)$.} 

\tablenotetext{b}{ $T_{\rm rad} \equiv$ Initial temperatures set from the equilibrium $T - \xi$ radiative heating-cooling function, 
i.e., the $T$ solution obtained by solving for ${\cal{L}} = 0$ in Eq.~(\ref{eq-netL}), 
assuming $\gamma = 5/3$.} 

\tablenotetext{c}{ Runs $24 - 28$ are started using the particle states from Run $23$ at a time $= 1.4$ Myr as the initial condition, 
and changing $L_{X}$ in each case.}

\end{deluxetable} 

%

\begin{deluxetable}{cccccccccc} 
\tablewidth{0pt} 
\tablecaption{Relevant Time Range for some Simulations} 
\tablehead{ 
\colhead{Run} & 
\colhead{$\dot{M}_{B} (\gamma_{\rm run}, \rho_{\infty}, T_{\infty})$ \tablenotemark{a}} & 
\colhead{$t_B$ \tablenotemark{b}} & 
\colhead{$r_{\rm out}$} & 
\colhead{$M_{\rm tot,IC}$} & 
\colhead{$M_{\rm tot,IC} / \dot{M}_{B}$ \tablenotemark{c}} & 
\colhead{$N$} & 
\colhead{$r_{\rm out} / (N^{\frac{1}{3}})$ \tablenotemark{d}} & 
\colhead{$t_{\rm steady}$ \tablenotemark{e}} 
\\ 
\colhead{No.} & 
\colhead{[g/s]} & 
\colhead{[$10^{4}$ yr]} & 
\colhead{[pc]} & 
\colhead{$[ M_{\odot} ]$} & 
\colhead{[$10^{4}$ yr]} & & 
\colhead{[pc]} & 
\colhead{[$10^{4}$ yr]} 
} 

\startdata 

$4$ & $4.6 \times 10^{27}$ & $0.79$ & $5$ & $1.81 \times 10^{6}$ & $2.48$ & $64^3$ & $0.078$ & $0.6$ \\ 

\hline \\ 

$5$ & $4.6 \times 10^{27}$ & $0.79$ & $10$ & $9.76 \times 10^{6}$ & $13.38$ & $64^3$ & $0.156$ & $1.6$ \\ 

$6$ & $4.6 \times 10^{27}$ & $0.79$ & $10$ & $9.76 \times 10^{6}$ & $13.38$ & $128^3$ & $0.078$ & $1.6$ \\ 

\hline \\ 

$7$ & $4.6 \times 10^{27}$ & $0.79$ & $20$ & $6.24 \times 10^{7}$ & $85.52$ & $128^3$ & $0.156$ & $4$ \\ 

\hline \\ 

$8$ & $4.6 \times 10^{27}$ & $0.79$ & $50$ & $8.48 \times 10^{8}$ & $1162.18$ & $128^3$ & $0.391$ & $8$ \\ 

\enddata 
\label{Table-Rout-Time} 

\tablenotetext{a}{ $\dot{M}_{B} =$ Bondi accretion rate for the simulation parameters: 
$\gamma_{\rm run} =  1.01$, $\rho_{\infty} =  10^{-19}$ g/cm$^3$, $T_{\infty} = 10^7$ K, $R_B = 3.0$ pc, $R_s = 1.5$ pc.} 

\tablenotetext{b}{ $t_B =$ Bondi time.} 
\tablenotetext{c}{ $M_{\rm tot,IC} / \dot{M}_{B} =$ Time within which all mass would be accreted in, assuming accretion at the Bondi rate.} 

\tablenotetext{d}{ $r_{\rm out} / (N^{1/3}) =$ Initial average inter-particle spacing, assuming uniform distribution of particles. 
It is a rough measure of simulation length resolution. 
Of course in the actual simulation the spatial resolution is given by the smoothing lengths of gas particles, 
which are dynamically determined at every timestep.} 

\tablenotetext{e}{ $t_{\rm steady} =$ Time for which simulation reproduces steady-state Bondi accretion.} 

\end{deluxetable} 


\clearpage

\begin{figure*} 
\centering 
\includegraphics[width = 0.8 \linewidth]{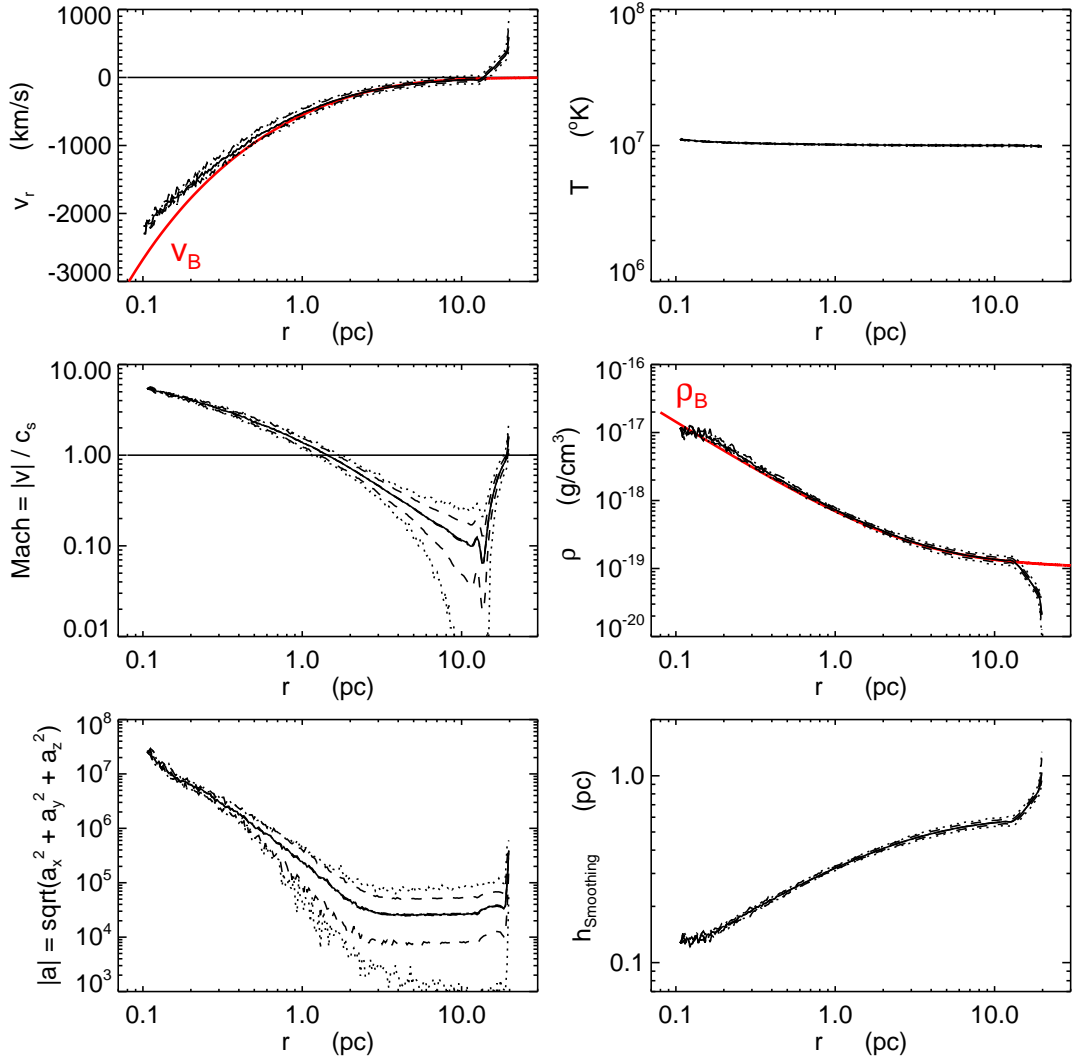}      
\caption{ 
Properties of particles in Run 7 as a function of radius, 
at simulation time $t = 2 t_B = 1.6 \times 10^{4}$ yr. 
The quantities plotted from top-left: 
radial velocity, temperature, Mach number, density, acceleration, and smoothing length.   
The solid line is the median value,     
the dashed lines are the $95$ percentile of the distribution lying below and above the median,    
and the dotted lines are the minimum and maximum values of the whole particle distribution at a given radius.   
The red curves in the top-left and middle-right panels show the Bondi solution for velocity and density.  
The horizontal straight lines in the top-left and middle-left panels indicate $v_r = 0$ and Mach $= 1$. 
} 
\label{fig-BondiScatterVsR} 
\end{figure*}

\begin{figure}
\centering 
\includegraphics[width = 0.6 \linewidth]{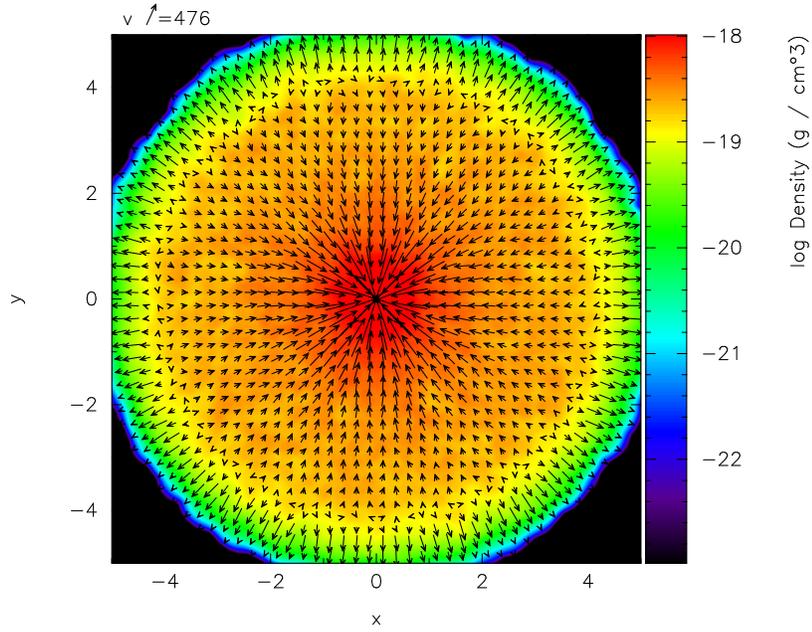}         
\caption{ 
Cross-section slice of gas density in the $x - y$ plane through $z = 0$ of Run 4 
at time $t = 0.25 t_B = 0.2 \times 10^{4}$ yr, overplotted with the velocity vector arrows. 
} 
\label{fig-Bondi-Splash} 
\end{figure} 

\begin{figure*}
\centering
\includegraphics[width = 0.8 \linewidth]{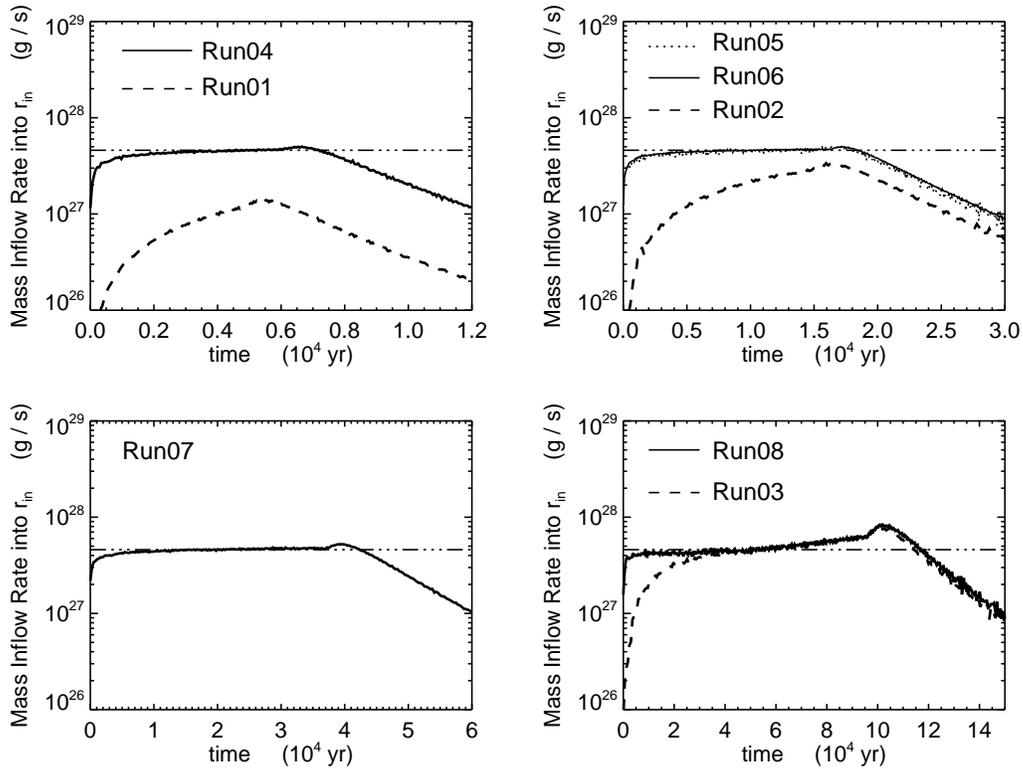}                   
\caption{ 
Mass inflow rate at the inner boundary as a function of time for the first eight runs in Table~\ref{Table-Bondi}. 
Each panel has a different outer radius: $r_{\rm out} = 5$ (top-left), $10$ (top-right), $20$ (bottom-left), $50$\,pc (bottom-right), 
as the time coverage becomes longer. 
The top-right panel shows different particle numbers: $N = 64^3$ (Run 05) and $128^3$ (Run 06) for the Bondi IC.  
In addition, the top row and bottom-right panels show the results of the Bondi IC (Runs 04, 05, 06, 08), 
together with the uniform IC runs (Runs 01, 02, 03). 
The Bondi mass accretion rate (marked as the 
dash-dot-dot-dot horizontal line in each panel) is reproduced for a limited time duration.
} 
\label{fig-Bondi-Mdot_Rout_N} 
\end{figure*} 

\begin{figure} 
\centering 
\includegraphics[width = 0.4 \linewidth]{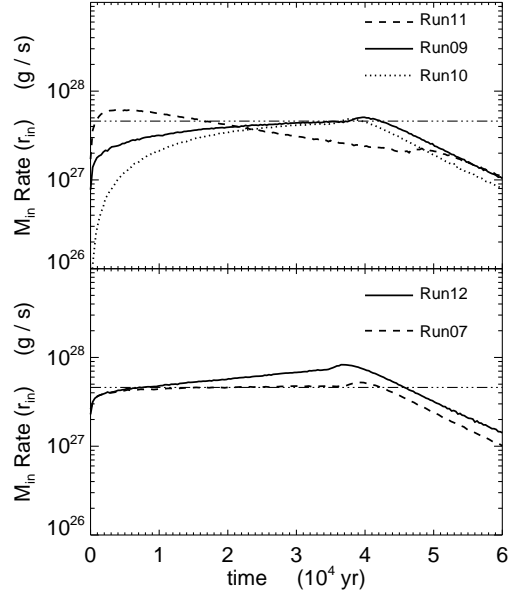}                    
\caption{ 
Mass inflow rate at the inner radius vs. time for the last four runs in Table~\ref{Table-Bondi}. 
The top panel shows results of different 
initial density profiles (see Table~\ref{Table-Bondi} and \S\ref{sec-results-Bondi-Inflow} for details). 
The bottom panel overplots two runs with the same Bondi IC, 
but including a galaxy bulge potential in one case (Run 12), and not in the other (Run 07). 
} 
\label{fig-Bondi-Mdot_IC_Bulge} 
\end{figure} 

\begin{figure*}
\centering
\includegraphics[width = 0.8 \linewidth]{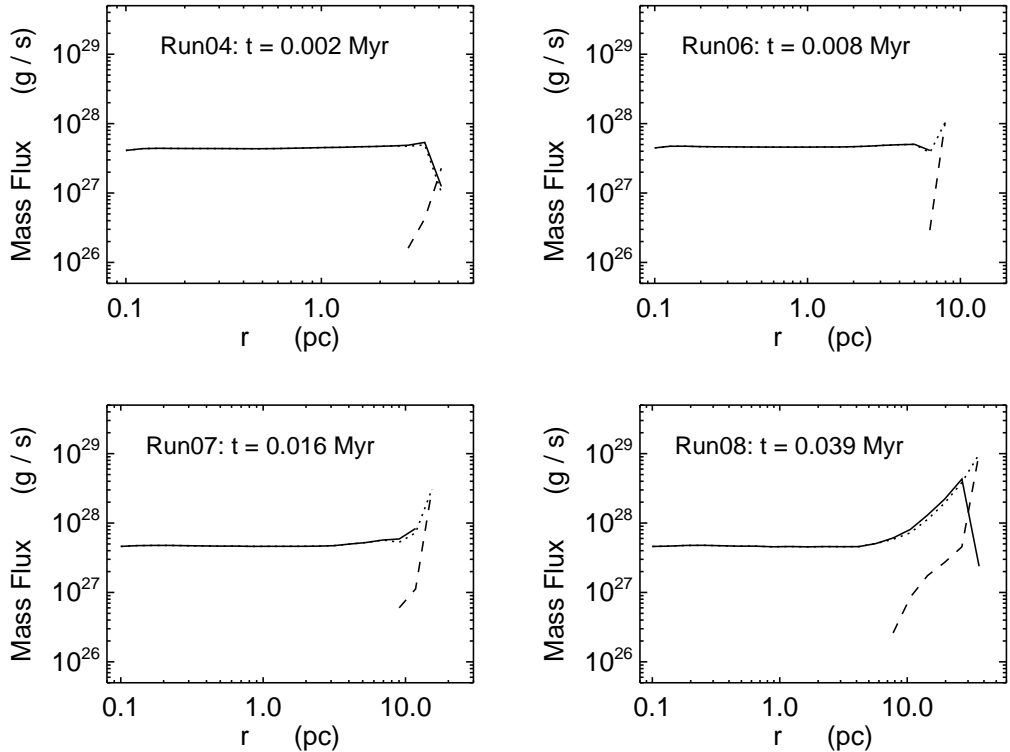}                
\caption{ 
Mass flux as a function of radius for four runs with different outer radius: 
$r_{\rm out} = 5$\,pc (top-left, Run 04), $10$\,pc (top-right, Run 06), $20$\,pc (bottom-left, Run 07), and $50$\,pc (bottom-right, Run 08). 
The inflow (solid line), outflow (dashed line), and net (dotted line) mass fluxes, 
as defined by Eq.~(\ref{eq-MassFlux2}), are separately plotted. 
The absolute value of a mass flux is shown whenever it is negative. 
$\dot{M}_{\rm in}$ is negative at all radii, and 
$\dot{M}_{\rm net}$ is negative in the inner parts where $| \dot{M}_{\rm in} | > | \dot{M}_{\rm out} |$. 
The simulation time of each run is noted in each panel. 
} 
\label{fig-Bondi-MassFlux} 
\end{figure*}

\begin{figure} 
\centering 
\includegraphics[width = 0.6 \linewidth]{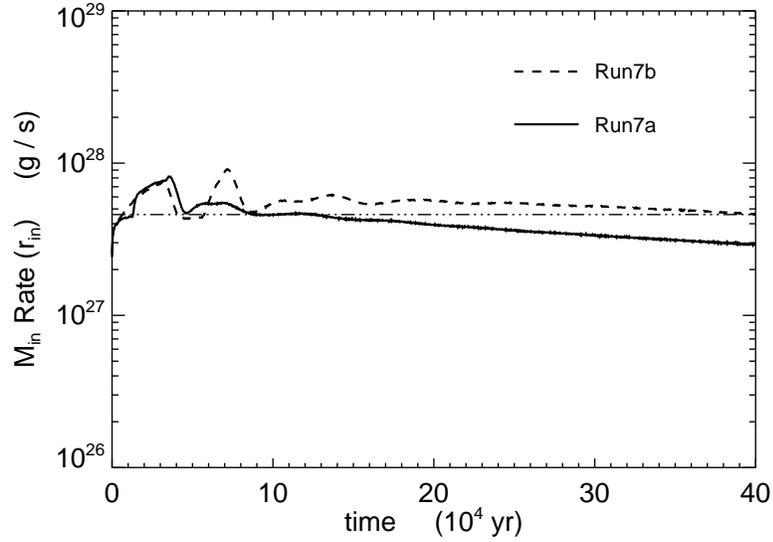}                    
\caption{ 
Mass inflow rate at the inner boundary vs. time for the two runs in Table~\ref{Table-Bondi} 
with different outer boundary conditions, which have substantially reduced outflow 
(see \S\ref{sec-results-Bondi-Discussion} for details). 
Inflow is maintained for durations $> 10$ times larger than in Run07 in Fig.~\ref{fig-Bondi-Mdot_Rout_N}. 
} 
\label{fig-Bondi-Mdot_OuterBC} 
\end{figure}

\begin{figure*} 
\centering 
\includegraphics[width = 0.8 \linewidth]{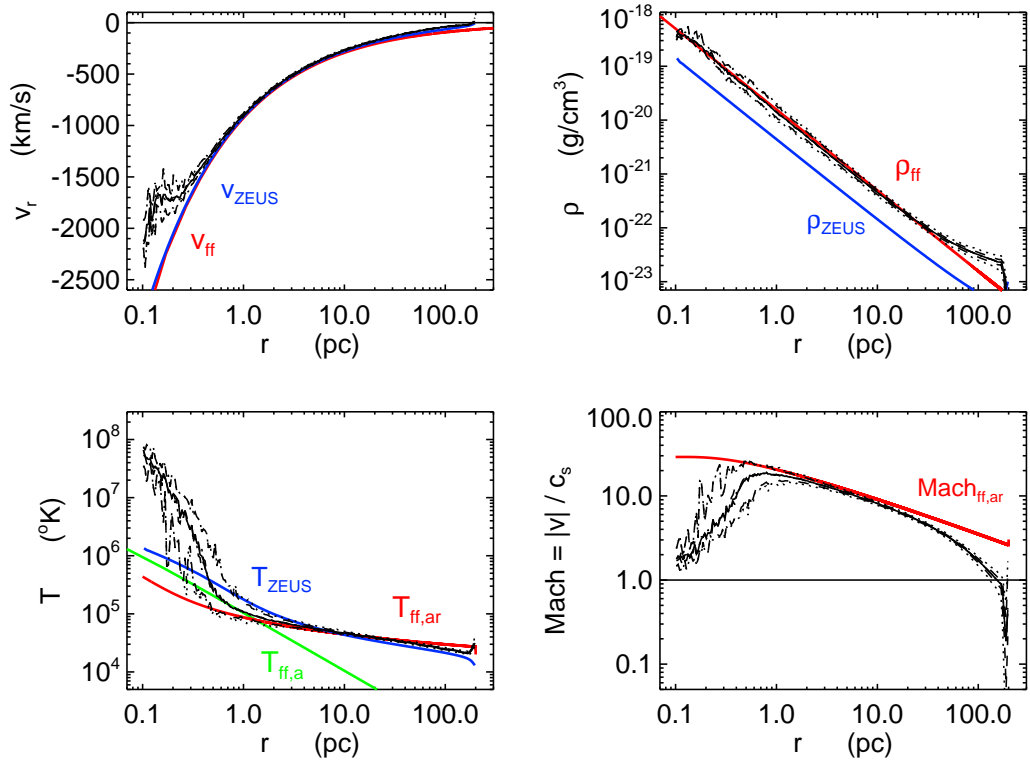}         
\caption{ 
Properties of particles in Run 23 (with heating and cooling) vs.~radius at $t = 1$ Myr. 
The quantities plotted are: 
radial velocity, density, temperature and Mach number.   
The solid line is the median value,      
the dashed lines are the $95$ percentile of the distribution lying below and above the median,     
and the dotted lines are the minimum and maximum values of the whole particle distribution at a given radius.    
The red curve in each panel shows the free-fall scaling of the corresponding quantity    
(discussed in \S\ref{sec-results-HC-CompareBondi}).    
The blue curve in the first three panels depict the corresponding ZEUS results:    
$v_{\rm ZEUS}$ and $v_{ff}$ are indistinguishable,    
$\rho_{\rm ZEUS}$ is at a constant offset from $\rho_{ff}$ because of different normalization.    
The green curve in the bottom-left panel indicate the free-fall temperature with only adiabatic processes.    
The horizontal straight lines in the top-left and bottom-right panels indicate $v_r = 0$ and Mach $= 1$.   
} 
\label{fig-HC-ScatterVsR} 
\end{figure*} 

\begin{figure*}
\centering
\includegraphics[width = 0.8 \linewidth]{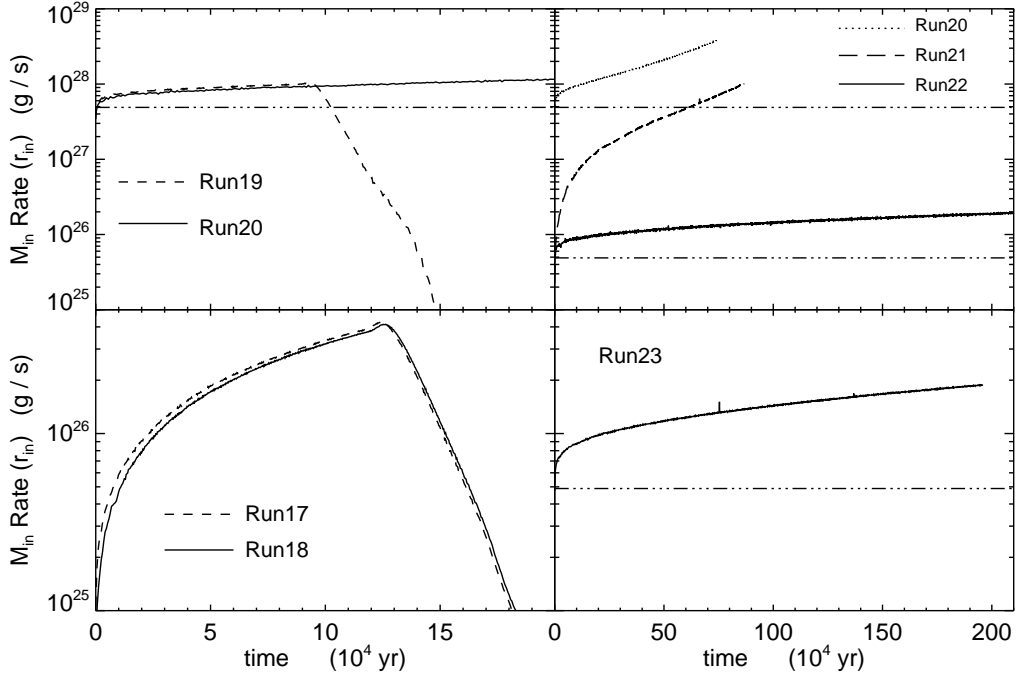}                  
\caption{ 
Mass inflow rate at the inner boundary vs.~time, 
for seven runs in Table~\ref{Table-HeatCool} with $T_{\rm init} = T_{\rm rad}$, i.e., the initial temperatures are in radiative equilibrium. 
Top-left panel shows that the accretion flow reaches almost steady-state with a larger outer radius. 
Top-right panel shows the effect of varying $T_{\infty}$ and $\rho_{\infty}$. 
Bottom-left panel shows that different $\gamma_{\rm init}$ has a negligible effect on the mass inflow. 
Note that the y-axis range in top and bottom panels are different. The details are discussed in \S\ref{sec-results-HC-Inflow}. 
} 
\label{fig-HC-Mdot_Rout200pc} 
\end{figure*} 

\begin{figure}
\centering
\includegraphics[width = 0.5 \linewidth]{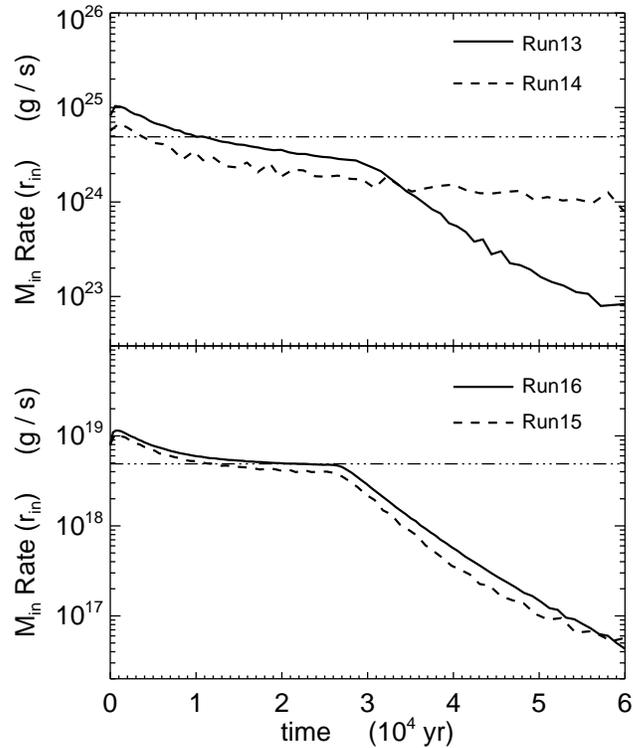}                
\caption{ 
Mass inflow rate across the inner radius as a function of time 
for the first four runs in Table~\ref{Table-HeatCool} with $T_{\rm init} = T_{\infty} = 10^7$ K. 
The top panel shows the effect of varying outer radius (details in \S\ref{sec-results-HC-Inflow}), and 
the bottom panel shows the results with different $L_{X}$ (\S\ref{sec-results-HC-Lx}). 
The 
dash-dotted horizontal line marks the corresponding Bondi accretion rate. 
} 
\label{fig-HC-Mdot_Rout_LX} 
\end{figure}

\begin{figure*}
\centering
\includegraphics[width = 0.9 \linewidth]{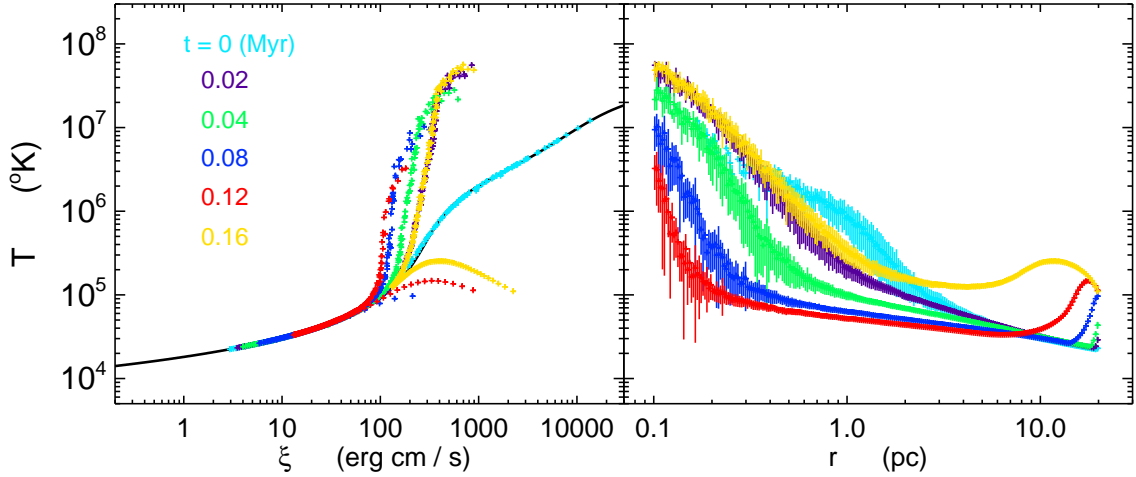}                       
\caption{ 
Time evolution of particles in Run 18 (whose mass inflow rate is shown in Figure~\ref{fig-HC-Mdot_Rout200pc}) in the $T - \xi$ and $T - r$ planes. 
Each color represents the particles at a certain simulation time, which are labeled in the left panel in Myrs.    
Note that high $\xi$ corresponds to small $r$.     
The left panel shows the median values of $T$ and $\xi$ at a given radius.     
The right panel shows the mean temperature and standard deviation around it plotted as vertical lines,      
indicating the particle distribution and scatter.   
} 
\label{fig-HC-Evol-T_Xi-T_r} 
\end{figure*}

\begin{figure}
\centering
\includegraphics[width = 0.7 \linewidth]{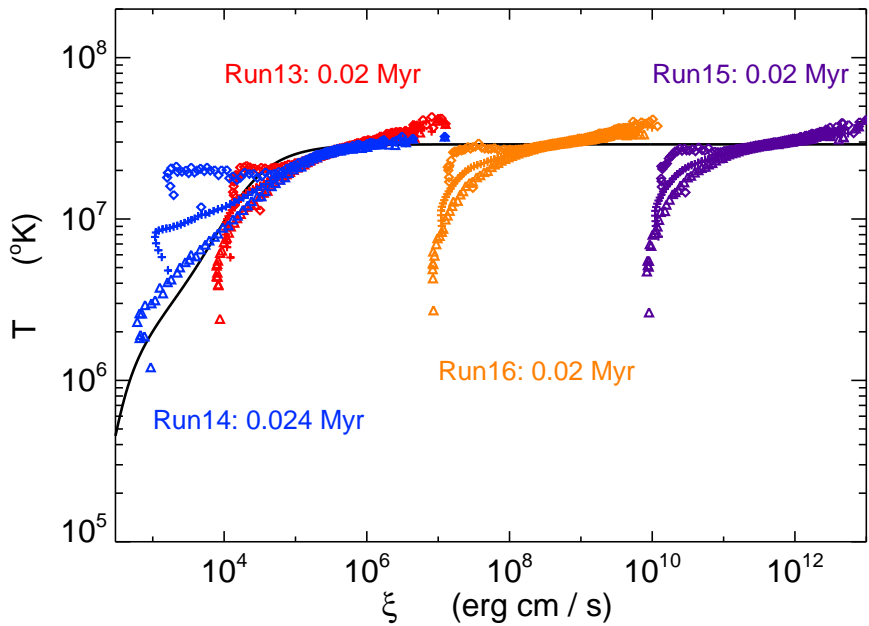}                   
\caption{ 
Temperature vs.~photoionization parameter of particles (details in \S\ref{sec-results-HC-Thermal}), at a simulation time as labeled for each run, 
for the first four runs in Table~\ref{Table-HeatCool} with $T_{\rm init} = 10^7$ K (whose mass inflow rates are shown in Figure~\ref{fig-HC-Mdot_Rout_LX}). 
The black line shows the radiative equilibrium $T - \xi$ curve    
for the implemented heating-cooling function (${\cal{L}} = 0$, Eq.~\ref{eq-netL}).      
The median values of $T$ and $\xi$ at a given radius are plotted as plus symbols,     
the maximum values as diamonds, and the minimum values as triangles.      
} 
\label{fig-HC-T_xi_Tinfinity} 
\end{figure}

\begin{figure}
\centering
\includegraphics[width = 0.8 \linewidth]{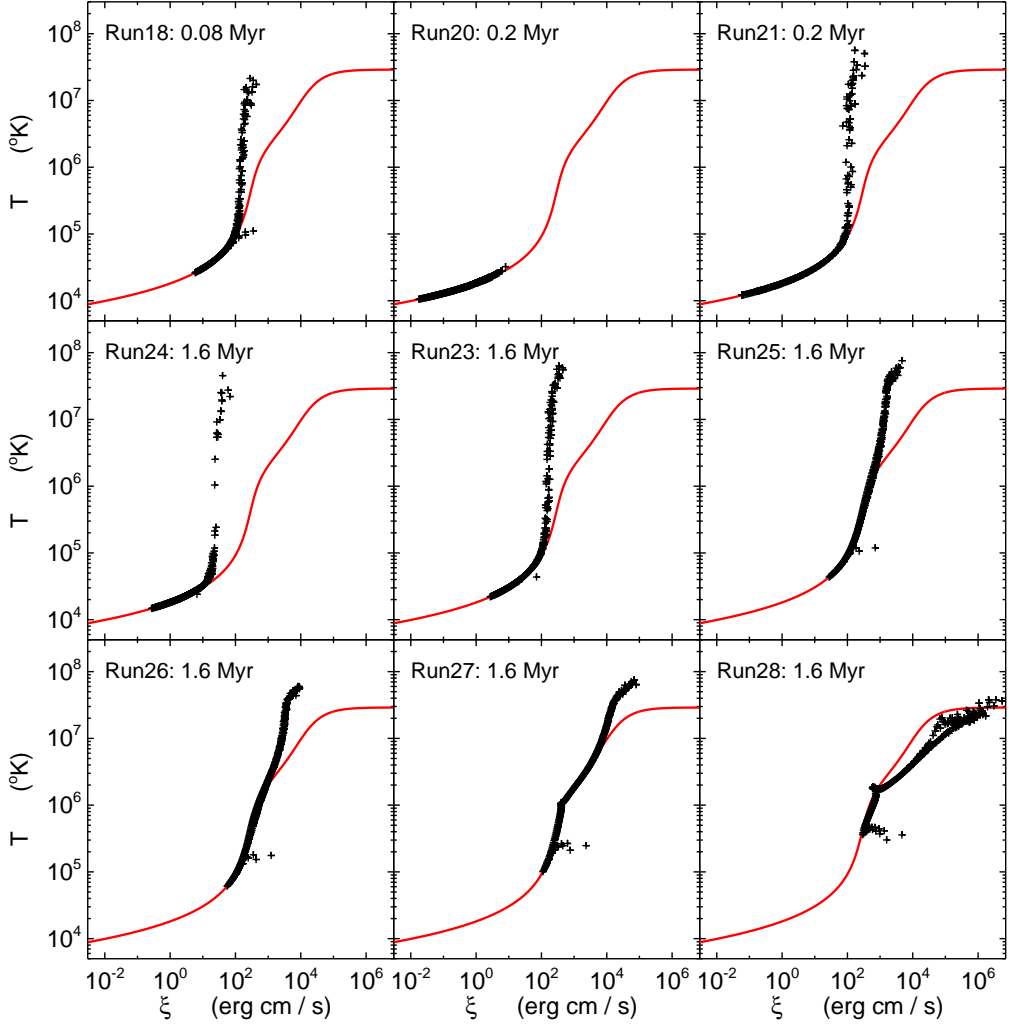}                    
\caption{ 
Temperature vs.~photoionization parameter of particles (discussed in \S\ref{sec-results-HC-Thermal}), 
for nine runs in Table~\ref{Table-HeatCool} with $T_{\rm init} = T_{\rm rad}$ 
(whose mass inflow rates are shown in Figures~\ref{fig-HC-Mdot_Rout200pc} and \ref{fig-HC-Mdot_LXray}).     
The median values of $T$ and $\xi$ at a given radius are plotted as the plus symbols.      
The radiative equilibrium $T - \xi$ curve is overplotted as the red line in each panel.    
} 
\label{fig-HC-T_xi_Tradiative} 
\end{figure}

\begin{figure*}
\centering
\includegraphics[width = 0.7 \linewidth]{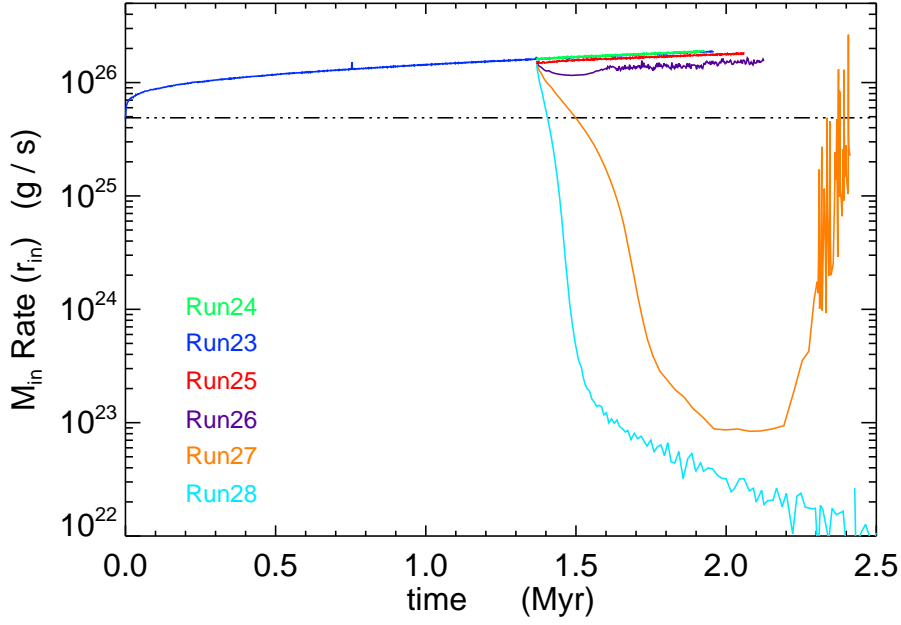}                    
\caption{ 
Mass inflow rate at the inner boundary vs.~time, 
for the last six runs in Table~\ref{Table-HeatCool} with varying $L_{X} / L_{\rm Edd}$ between $5 \times 10^{-5}$ and $5 \times 10^{-2}$. 
There is a significant drop in the mass inflow rate when $L_{X}$ is increased from $0.01$ to $0.02$, 
which is associated with the development of an outflow. 
The details are discussed in \S\ref{sec-results-HC-Lx}. 
} 
\label{fig-HC-Mdot_LXray} 
\end{figure*} 

\begin{figure*}
\centering
\includegraphics[width = 1.0 \linewidth]{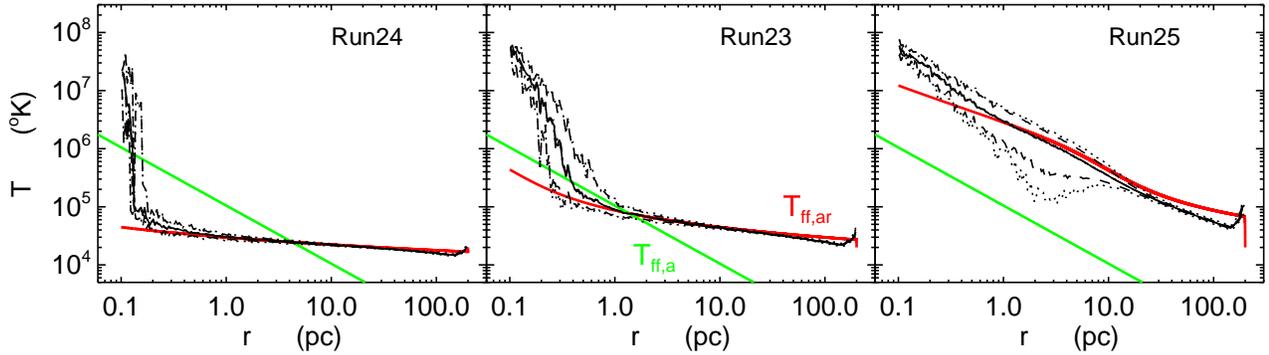}                          
\caption{ 
Radial temperature profiles of particles in three runs with different X-ray luminosity: 
$L_{X} / L_{\rm Edd} = 5 \times 10^{-5}$ (Run 24), $5 \times 10^{-4}$ (Run 23), $5 \times 10^{-3}$ (Run 25), at a simulation time $t = 1.6$ Myr.    
The solid line is the median value,       
the dashed lines are the $95$ percentile of the distribution lying below and above the median,      
and the dotted lines are the minimum and maximum values of the whole particle distribution at a given radius.     
The overplotted colored lines are the solutions of the steady-state internal energy equation, assuming free-fall scalings of the velocity and density:    
considering both adiabatic and radiative terms $T_{ff,ar}$ (\S\ref{sec-appenix}) is plotted as the red curve,  
and considering only adiabatic terms $T_{ff, a}$ (Eq.~\ref{eq-T_ffa}) is plotted as the green curve.  
The details are discussed in \S\ref{sec-results-HC-AV}. 
} 
\label{fig-HC-T_r_3Lx} 
\end{figure*}

\begin{figure*} 
\centering 
\includegraphics[width = 0.8 \linewidth]{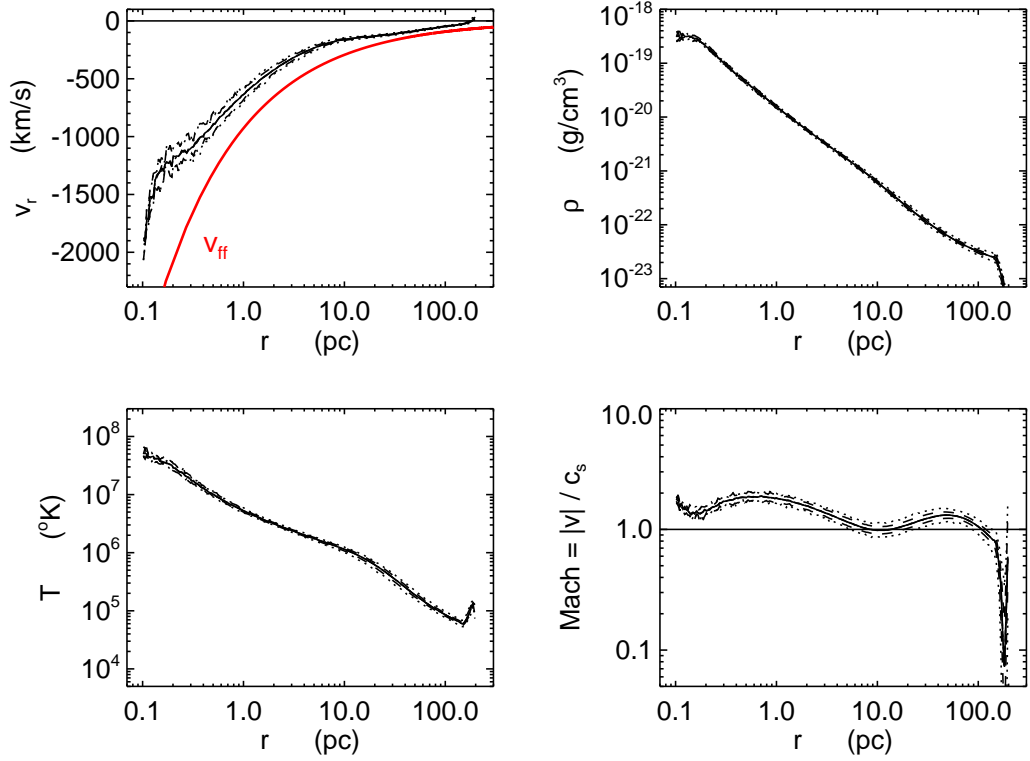}                  
\caption{ 
Properties of particles in Run 26 ($L_{X} / L_{\rm Edd} = 0.01$) vs.~radius at $t = 1.5$ Myr, 
using a similar format as in Figure~\ref{fig-HC-ScatterVsR}. 
} 
\label{fig-HC-ScatterVsR-Lx1e-2} 
\end{figure*} 

\begin{figure*} 
\centering 
\includegraphics[width = 0.8 \linewidth]{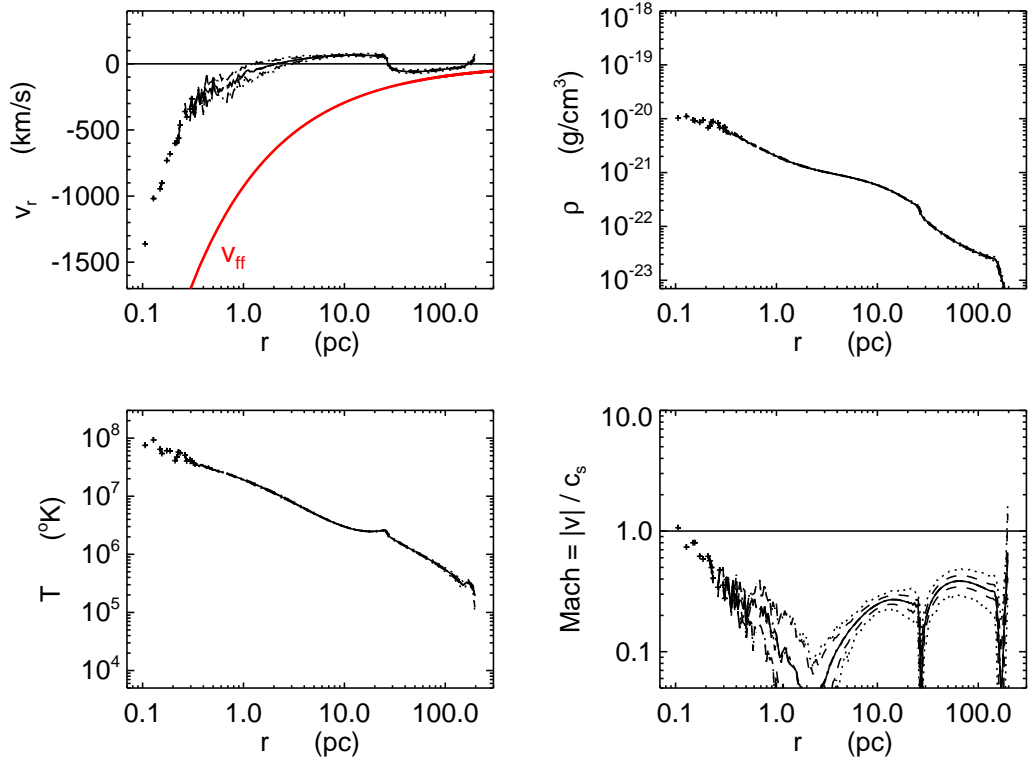}                 
\caption{ 
Properties of particles in Run 28 ($L_{X} / L_{\rm Edd} = 0.05$) vs.~radius at $t = 1.5$ Myr, 
using a similar format as in Figure~\ref{fig-HC-ScatterVsR}.     
The median values at few inner radial bins are plotted as plus symbols.   
} 
\label{fig-HC-ScatterVsR-Lx5e-2} 
\end{figure*}

\end{onecolumn}

\end{document}